\documentclass[12pt,oneside]{amsart}
\usepackage{amssymb,mathrsfs}
\usepackage{tabularx}
\usepackage{epsfig}
\usepackage[hypertex,linkcolor=red]{hyperref}
\usepackage{fancybox,fancyhdr}
\usepackage{txfonts}
\usepackage[matrix,frame,arrow]{xy}
\usepackage{amsmath}

\newcommand{\qw}[1][-1]{\ar @{-} [0,#1]}
\newcommand{\gate}[1]{*{\xy *+<.6em>{#1};p\save+LU;+RU **\dir{-}\restore\save+RU;+RD **\dir{-}\restore\save+RD;+LD **\dir{-}\restore\POS+LD;+LU **\dir{-}\endxy} \qw}
\newcommand{\measureD}[1]{*{\xy*+=+<.5em>{\vphantom{\rule{0em}{.1em}#1}}*\cir{r_l};p\save*!R{#1} \restore\save+UC;+UC-<.5em,0em>*!R{\hphantom{#1}}+L **\dir{-} \restore\save+DC;+DC-<.5em,0em>*!R{\hphantom{#1}}+L **\dir{-} \restore\POS+UC-<.5em,0em>*!R{\hphantom{#1}}+L;+DC-<.5em,0em>*!R{\hphantom{#1}}+L **\dir{-} \endxy} \qw}
\newcommand{\multimeasureD}[2]{*+<1em,.9em>{\hphantom{#2}}\save[0,0].[#1,0];p\save !C *{#2},p+LU+<0em,0em>;+RU+<-.8em,0em> **\dir{-}\restore\save +LD;+LU **\dir{-}\restore\save +LD;+RD-<.8em,0em> **\dir{-} \restore\save +RD+<0em,.8em>;+RU-<0em,.8em> **\dir{-} \restore \POS !UR*!UR{\cir<.9em>{r_d}};!DR*!DR{\cir<.9em>{d_l}}\restore \qw}
\newcommand{\multigate}[2]{*+<1em,.9em>{\hphantom{#2}} \qw \POS[0,0].[#1,0];p !C *{#2},p \save+LU;+RU **\dir{-}\restore\save+RU;+RD **\dir{-}\restore\save+RD;+LD **\dir{-}\restore\save+LD;+LU **\dir{-}\restore}
\newcommand{\ghost}[1]{*+<1em,.9em>{\hphantom{#1}} \qw}
\newcommand{\Qcircuit}[1][0em]{\xymatrix @*[o] @*=<#1>}
\newcommand{\pureghost}[1]{*+<1em,.9em>{\hphantom{#1}}}
\newcommand{\multiprepareC}[2]{*+<1em,.9em>{\hphantom{#2}}\save[0,0].[#1,0];p\save !C
  *{#2},p+RU+<0em,0em>;+LU+<+.8em,0em> **\dir{-}\restore\save +RD;+RU **\dir{-}\restore\save
  +RD;+LD+<.8em,0em> **\dir{-} \restore\save +LD+<0em,.8em>;+LU-<0em,.8em> **\dir{-} \restore \POS
  !UL*!UL{\cir<.9em>{u_r}};!DL*!DL{\cir<.9em>{l_u}}\restore}
\newcommand{\prepareC}[1]{*{\xy*+=+<.5em>{\vphantom{#1\rule{0em}{.1em}}}*\cir{l^r};p\save*!L{#1} \restore\save+UC;+UC+<.5em,0em>*!L{\hphantom{#1}}+R **\dir{-} \restore\save+DC;+DC+<.5em,0em>*!L{\hphantom{#1}}+R **\dir{-} \restore\POS+UC+<.5em,0em>*!L{\hphantom{#1}}+R;+DC+<.5em,0em>*!L{\hphantom{#1}}+R **\dir{-} \endxy}}
\newcommand{\poloFantasmaCn}[1]{{{}^{#1}_{\phantom{#1}}}}

\def\eg{{e.~g.} }\def\ie{{i.~e.} }
\def\paragraph#1{\medskip\par\noindent{\bf #1}}

\def\Tr{\operatorname{Tr}}
\def\Reals{\mathbb R}\def\qed{$\,\blacksquare$\par}
\renewcommand{\geq}{\geqslant}\renewcommand{\leq}{\leqslant}

\def\mat#1{{\boldsymbol{#1}}}
\def\transp#1{{#1}^{T}}

\def\Her{\set{Her}}
\def\openone{\leavevmode\hbox{1\kern-3.8pt 1}}% 
\def\set#1{{\sf #1}}\def\alg#1{{\mathcal #1}}\def\aA{\alg{A}}
\def\grp#1{{\mathbf #1}}

\def\Span{\set{Span}}
\def\Prob{\mathbb{P}}

\def\Erays{\set{Erays}}

\def\<{\langle}\def\>{\rangle}\def\kk{\rangle\!\!\rangle}
\def\bb{\langle\!\!\langle}

\def\Bnd#1{\set{Bnd(#1)}}\def\Bndp#1{\set{Bnd}_+(#1)}
\def\Lin#1{\set{Lin(#1)}}

\def\sH{\set{H}}
\def\sA{\set{A}}\def\sS{\set{S}}
\def\K#1{\left|#1\right)\!}
\def\B#1{\left(#1\right|}
\def\SC#1#2{\left(#1\left|\right.\!#2\right)\!}

\def\trnsfrm#1{\mathscr #1}\def\lntrnsfrm#1{\mathrm #1} \def\tA{\trnsfrm A}\def\tB{\trnsfrm
  B}\def\tC{\trnsfrm C}\def\tD{\trnsfrm D} \def\tS{\trnsfrm S}\def\tG{\trnsfrm G} \def\tI{\trnsfrm I}\def\tU{\trnsfrm U} \def\tF{\trnsfrm F}\def\tE{\trnsfrm E} \def\tR{\trnsfrm R}
\def\rA{\lntrnsfrm
  A}\def\rB{\lntrnsfrm B}\def\rC{\lntrnsfrm C} \def\rD{\lntrnsfrm D}\def\rE{\lntrnsfrm
  E}\def\rF{{\lntrnsfrm F}} \def\rG{{\lntrnsfrm G}}\def\rI{\lntrnsfrm I}\def\rH{\lntrnsfrm H}
\def\rL{{\lntrnsfrm L}}\def\rM{\lntrnsfrm M}\def\rN{\lntrnsfrm N}\def\rO{{\lntrnsfrm O}}
\def\rP{\lntrnsfrm P}
   \def\rX{{\lntrnsfrm X}} \def\rY{{\lntrnsfrm Y}} 
\def\AA{\mathbb
  A}  
\def\Stset{{\mathfrak S}}
\def\Trnset{{\mathfrak T}}
\def\Cntset{{\mathfrak E}}

\def\dim{\operatorname{dim}}

 \def\qA{\mathcal A}\def\qD{\mathcal D}  \def\qI{\mathcal
  I}

\def\eg{{\em e.g.} }\def\ie{{\em i.e.} }
\def\<{\langle}\def\>{\rangle}\def\kk{\rangle\!\rangle}
\def\bb{\langle\!\langle}
\def\Her{\set{Her}}
\def\qnt{\Upsilon}
\def\mat#1{{\boldsymbol{#1}}}
\def\Prob{\mathbb{P}}
\def\<{\langle}\def\>{\rangle}\def\Reals{\mathbb R}
\def\set#1{{\sf #1}}
\def\grp#1{{\mathbf #1}}
\def\transp#1{{#1}^{\, t}}
\def\Stset{{\mathfrak S}}\def\Trnset{{\mathfrak T}}\def\Cntset{{\mathfrak E}} 
\def\Span{\set{Span}}\def\Bnd#1{\set{Lin(#1)}}\def\Bndp#1{\set{Lin}_+(#1)}
\def\Erays{\set{Erays}}\def\Extr{\operatorname{Extr}}
\def\sH{\set{H}}\def\qed{$\,\blacksquare$\par}
\def\eff{{\rm eff}}\renewcommand{\geq}{\geqslant}\renewcommand{\leq}{\leqslant}
\def\Bnd#1{\set{Lin(#1)}}\def\Lin#1{\set{Lin(#1)}}
\def\Postulate#1#2#3{\medskip\par\noindent{\bf Postulate #1: #2.$\;$}{\em #3}\medskip\par}
\def\Definition#1{\medskip\par\noindent{\bf Definition:$\;$}{#1}\medskip\par}

\newtheorem{lemma}{Lemma}\newtheorem{theorem}{Theorem}[section]\newtheorem{corollary}{Corollary}[section]
\newtheorem{proposition}{Proposition}[section]\newtheorem{observation}{Observation}[section]

\def\Proof{\medskip\par\noindent{\bf Proof. }}
\def\trnsfrm#1{\mathscr #1}\def\lntrnsfrm#1{\mathrm #1}
\def\aA{\alg{A}}\def\tA{\trnsfrm A}\def\tB{\trnsfrm B}
\def\tC{\trnsfrm C}\def\tD{\trnsfrm D}\def\tI{\trnsfrm I}
\def\tS{\trnsfrm S}\def\tU{\trnsfrm U}
\def\tR{\trnsfrm R}
\def\AA{\mathbb A}

\def\qA{\mathcal A}\def\qI{\mathcal I}\def\qD{\mathcal D}
\def\sH{\set{H}}\def\sS{\set{S}}
\def\Stset{{\mathfrak S}}\def\alg#1{{\mathcal #1}}
\def\Tr{\operatorname{Tr}}
\def\b#1#2#3#4#5#6#7#8#9{\begin{matrix}\\ \begin{bmatrix}#1& #2& #3\\ #4& #5& #6\\ #7& #8& #9\end{bmatrix}\\{} \end{matrix}}
\begin{document}
\title[Testing axioms for Quantum Mechanics on toy-theories]{Testing axioms for Quantum Mechanics\\ on Probabilistic toy-theories}
\author{Giacomo Mauro D'Ariano\\Alessandro Tosini}
%% Abstract %%
\begin{abstract}
  In Ref. \cite{DAriano:2008p3785} one of the authors proposed postulates for axiomatizing Quantum
  Mechanics as a {\em fair operational framework}, namely regarding the theory as a set of rules
  that allow the experimenter to predict future events on the basis of suitable tests, having local
  control and low experimental complexity. In addition to causality, the following postulates have
  been considered: PFAITH (existence of a pure preparationally faithful state), and FAITHE
  (existence of a faithful effect). These postulates have exhibited an unexpected theoretical power,
  excluding all known nonquantum probabilistic theories.  Later in Ref.  \cite{Chiribella:2009unp}
  in addition to causality and PFAITH, local discriminability and PURIFY (purifiability of all
  states) have been postulated, narrowing the probabilistic theory to something very close to
  Quantum Mechanics.  In the present paper we test the above postulates on some nonquantum
  probabilistic models.  The first model, {\em the two-box world} is an extension of the
  Popescu-Rohrlich model\cite{Rohrlich:1995p1940}, which achieves the greatest violation of the CHSH
  inequality compatible with the no-signaling principle.  The second model {\em the two-clock world}
  is actually a full class of models, all having a disk as convex set of states for the local
  system.  One of them corresponds to the {\em the two-rebit world}, namely qubits with real Hilbert
  space. The third model---{\em the spin-factor}---is a sort of $n$-dimensional generalization of
  the clock.  Finally the last model is {\em the classical probabilistic theory}.  We see how each
  model violates some of the proposed postulates, when and how teleportation can be achieved, and we
  analyze other interesting connections between these postulate violations, along with deep
  relations between the local and the non-local structures of the probabilistic theory.
\end{abstract}

\maketitle
%% Introduction %%
\section{Introduction}
Quantum Mechanics is still laking a foundation. The Lorentz transformations suffered the same
problem before the discovery of special relativity, and an analogous principle of ``quantumness''
has not been found yet. If one considers the theoretical power of special relativity in the ensuing
research, one definitely ought to put the principle of quantumness at the highest research priority.
\par In the recent article \cite{DAriano:2008p3785} one of the authors proposed to axiomatize
Quantum Mechanics as a {\em fair operational framework}, namely regarding the theory as of rules
that allow the experimenter to predict future events on the basis of suitable tests, having local
control and low experimental complexity. In addition to causality, the following postulates have
been considered: PFAITH (existence of a pure preparationally faithful state), and FAITHE (existence
of a faithful effect). These postulates have exhibited an unexpected theoretical power, excluding
all known nonquantum probabilistic theories, such as PR-boxes \cite{Rohrlich:1995p1940}, {\em
  rebits} \cite{Wootters}, etc. The two postulate alone are however not sufficient to derive Quantum
Mechanics, and other potential postulates of the same nature have been considered, such as FAITHE:
(existence of a faithful effect), and SUPERFAITH (existence of a pure preparationally state which
used in many copies also provides a $2n$-partite preparationally faithful states). More recently in
Ref.  \cite{Chiribella:2009unp} a more extensive axiomatic approach has been used, and in addition to
NSF and PFAITH, postulate LDISCR (local discriminability) and PURIFY (purifiability of all
  states, uniquely up to reversible channes on the purifying system) have been considered. These
postulates make the probabilistic framework much closer to Quantum Mechanics, with teleportation,
error correction, dilation theorems, no cloning, and no bit commitment among its corollaries.

In the present paper we test the above postulates on the available probabilistic models different
from Quantum Mechanics. The first model, {\em the two-box world}, is an extension of the
Popescu-Rohrlich model \cite{Rohrlich:1995p1940}, which achieves the greatest violation of the CHSH
inequality compatible with the no-signaling principle.  The second model, {\em the two-clock world},
is actually a full class of models, all having a disk as convex set of states for the local system.
These models allow purification of all its mixed states, but the purification is not unique up to
reversible channels on the purifying system, as PURIFY requires.  One of the models of this class is
indeed the {\em the two-rebit world}, namely qbits with real Hilbert space. This model violates the
local observability principle, namely the possibility of discriminating joint states by local
measurements. The third model---{\em the spin-factor}---is a sort of $n$-dimensional generalization
of the clock. Here we show that the only dimension $n=3$ allows teleportation, and, indeed, in such
case the theory is the \emph{qubit}.  Finally the last model is {\em the classical probabilistic
  theory}. We see how each model violates some of the proposed postulates, when and how
teleportation can be achieved, along with interesting connections with the violations of postulates
and deep relations between the local and the non-local structures of the probabilistic theory.

\par The world of probabilistic theories is still largely unexplored, and we still have poor
intuition. Mostly our intuition is biased by our familiarity with Quantum Mechanics, and it is easy
to mistakenly assume quantum features as general properties of probabilistic theories.  This is also
a consequence of the absence of available alternative probabilistic models to test the new postulates.
This is the main motivation for the present paper, where some concrete probabilistic models
alternative to Quantum Mechanics are constructed and analyzed.
\section{Short review on probabilistic operational theories}  
\subsection{The operational framework.}
The primitive notion of our framework is the notion {\bf test}. A test is made of the following
ingredients: a) a complete collection of {\bf outcomes}, b) input {\bf systems}, c)
output systems. It is represented in form of a box, as follows

\medskip\centerline{\large
$\Qcircuit @C=1em @R=.7em @! R {
&\qw\poloFantasmaCn{\rA_1}&\multigate{1}{\{\tA_i\}}&\qw\poloFantasmaCn{\rB_1}&\qw\\
&\qw\poloFantasmaCn{\rA_2}&\ghost{\{\tA_i\}}&\qw\poloFantasmaCn{\rB_2}&\qw}$\qquad
$\Qcircuit @C=1em @R=.7em @! R {
&\qw\poloFantasmaCn{\rA_1}&\multigate{1}{\tA}&\qw\poloFantasmaCn{\rB_1}&\qw\\
&\qw\poloFantasmaCn{\rA_2}&\ghost{\tA}&\qw\poloFantasmaCn{\rB_2}&\qw}$
}\medskip

The left wires represent the input systems, the right wires the output systems, and $\{\tA_i\}$ the
collection of outcomes.  Very often it is convenient to represent not the complete test, but just a
single outcome $\tA_i$, or, more generally a subset $\tA\subset\{\tA_i\}$ of the collection of
possible outcomes, \ie what is called {\bf event}.  The number of wires at the input and at the
output can vary, and one can also have no wire at the input and/or at the input. We can regard the
test in many different ways, depending on our needs and context. A test can be a man-made
apparatus---such as a Stern-Gerlach setup or a beam splitter---or a nature-made
``phenomenon''---such as a physical interaction between different particles in some space-time
region. The set of events of a test is closed under union, intersection, and complementation, thus
making a Boolean algebra. The {\bf union} $\tA\cup\tB$ of two events $\tA$ or $\tB$ is the event in
which either $\tA$ or $\tB$ occurred, but it is unknown which one.  This operation is also called
{\bf coarse-graining}.  Reversely, a {\bf refinement} of an event $\tA$ is a set of events
$\{\tA_i\}$ occurring in some test such that $\tA=\cup_i \tA_i$.  Generally an event has different
refinements, depending on the test, and is not refinable within some test. We will call an event
that is unrefinable within any test {\bf atomic event}.

\paragraph{Connecting the test in a network.}
The natural place for a test/event will be inside a network of other tests/events, and to understand
the origin of the box representation and the intimate meaning of the test/event you should imagine
it actually connected to other tests/events in a circuit, \eg as follows
$$
\Qcircuit @C=1em @R=.7em @! R {
\multiprepareC{3}{\Psi}&\qw\poloFantasmaCn{\rA}&\multigate{1}{\tA}&\qw\poloFantasmaCn{\rB}&\gate{\tC}&\qw\poloFantasmaCn{\rC}&\multigate{1}{\tE}&\qw\poloFantasmaCn{\rD}&\multimeasureD{2}{\tG}\\
\pureghost{\Psi}&\qw\poloFantasmaCn{\rE}&\ghost{\tA}&\qw\poloFantasmaCn{\rF}&\multigate{1}{\tD}&\qw\poloFantasmaCn{\rG}&\ghost{\tE}&&\pureghost{\tG}\\
\pureghost{\Psi}&\qw\poloFantasmaCn{\rH}&\multigate{1}{\tB}&\qw\poloFantasmaCn{\rL}&\ghost{\tD}&\qw\poloFantasmaCn{\rM}&\multigate{1}{\tF}&\qw\poloFantasmaCn{\rN}&\ghost{\tG}\\
\pureghost{\Psi}&\qw\poloFantasmaCn{\rO}&\ghost{\tB}&\qw\poloFantasmaCn{\rP}&\qw&\qw &\ghost{\tF}\\
}$$
The different letters $\rA,\rB,\rC,\ldots$ labeling the wires precisely denote different
``types of system'', whose meaning comes from the following rules:
\paragraph{Connectivity rules:} (1) we can connect only an input wire of a box with an output wire
of another box, (2) we can connect only wires with the same label, (3) loops are forbidden.

The fact that there are no closed loops gives to the circuit the structure of a {\bf directed
  acyclic graph} (DAG). In the typical graph representation {\bf vertices} correspond to operations,
and {\bf edges} to wires. The circuit and the graph representations are exactly equivalent, once one
looks at a vertex as a ``box'' with inputs and outputs, as follows

$$
\begin{matrix}
\Qcircuit @C=1em @R=.7em @! R {&\multigate{2}{~^{}}&\qw\\
&\pureghost{~^{}}&\qw\\
&\ghost{~^{}}&\qw
}
\end{matrix}\hskip 1cm\Longleftrightarrow\hskip 1cm
\begin{matrix}
\epsfig{file=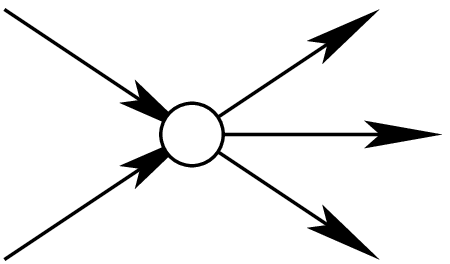,width=2cm}
\end{matrix}
$$

Ultimately the wires have only the function of ruling the way in which a box can be connected to
other boxes. Thus {\em systems are just a representation of the causal connections between different
  events}. The fact that there are no closed loops corresponds to the requirement that the
test/event is one-use only, whence each box in the circuit represents events that happen only once.
Moreover, we must keep in mind that the probability of the event is independent on the test which it
belongs, in the sense that if we have another test that contains the same event, this will have the
same probability (keeping the rest of the network fixed). The fact that the probability depends only
on the event and not on the test legitimates our use of networks made of single-event boxes, where
on each box we don't need to specify the test. In the following, we will denote the set of events
from system $\rA$ to system $\rB$ as $\Trnset(\rA,\rB)$, and use the abbreviation
$\Trnset(\rA):=\Trnset(\rA,\rB)$.

\paragraph{The trivial system.} Among the different kinds of systems, we consider a special one called
{\bf trivial system}, denoted by $\rI$. In the circuit it will be represented by no wire, but
instead we will draw the corresponding side of the operation box convexly rounded, namely as follows
$$\Qcircuit @C=1em @R=.7em @! R
  {\prepareC{\omega}&\qw\poloFantasmaCn{\rA}&\qw}\Longleftrightarrow\Qcircuit 
@C=1em @R=.7em @! R { &\qw \poloFantasmaCn{\rI}&\gate{\omega}&\qw\poloFantasmaCn{\rA}&\qw}
\qquad \Qcircuit @C=1em @R=.7em @! R{&\qw\poloFantasmaCn{\rA}&\measureD{a}}\Longleftrightarrow 
\Qcircuit @C=1em @R=.7em @! R {&\qw\poloFantasmaCn{\rA}&\gate{a}&\qw\poloFantasmaCn{\rI}&\qw}\;. $$

\paragraph{Building up the network formally.} One can build up the network using formal rules as in
Ref. \cite{Chiribella:2009unp}, making connection in parallel, in sequence, declaring commutativity
of parallel composition, etc. This construction is mathematically equivalent to the construction of
a \emph{symmetric strict monoidal category}, and poses a strong bridge with the research line of
Coecke and Abramsky \cite{abracoecke}. We also must keep in mind that there are no constraints for
disconnected parts of the network, namely they can be arranged freely as long as they are
disconnected (this for example would not be true for a quaternionic quantum network). Finally, we
will also consider \emph{randomized tests}, where one can choose a different test depending on the
outcome of a previous one. Such tests are provably feasible in causal theories.

\subsection{The operational probabilistic theory}

If you now want to make predictions about the occurrence probability of events based on your current
knowledge, then you need a ``theory'' that assign probabilities to different
events:\footnote{Probabilities in the network can be introduced in a easy intuitive way, or in a
  more axiomatic way as Ref. \cite{Chiribella:2009unp}.} {\em An {\bf operational theory} is
  specified by a collection of systems, closed under parallel composition, and by a collection of
  tests, closed under parallel/sequential composition and under randomization. The operational
  theory is \emph{probabilistic} if every test from the trivial system to the trivial system is
  associated to a probability distribution of outcomes.}

Therefore a probabilistic theory provides us with the joint probabilities for all possible events in
each box for any closed network, namely which has no input and no output system. The probability
itself will be conveniently represented by the corresponding network of events.  One is seldom
interested in full joint probabilities, but, more often, in the joint probability of events in some
given tests in the network, irrespective of events in all other tests. This will correspond to
marginalize over the other tests. We will see how the evaluation of probabilities will be greatly
simplified by the causality assumption and by the use of conditional states.

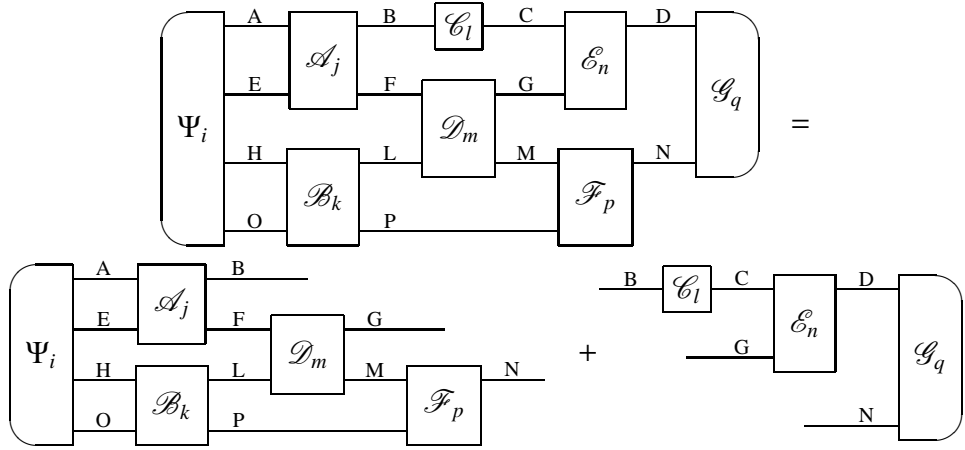
\begin{figure}[t]
$$\begin{matrix}\Qcircuit @C=1em @R=.7em @! R {
\multiprepareC{3}{\Psi_i}&\qw\poloFantasmaCn{\rA}&\multigate{1}{\tA_j}&\qw\poloFantasmaCn{\rB}&\gate{\tC_l}&\qw\poloFantasmaCn{\rC}&\multigate{1}{\tE_n}&\qw\poloFantasmaCn{\rD}&\multimeasureD{2}{\tG_q}\\
\pureghost{\Psi_i}&\qw\poloFantasmaCn{\rE}&\ghost{\tA_j}&\qw\poloFantasmaCn{\rF}&\multigate{1}{\tD_m}&\qw\poloFantasmaCn{\rG}&\ghost{\tE_n}&&\pureghost{\tG_q}\\
\pureghost{\Psi_i}&\qw\poloFantasmaCn{\rH}&\multigate{1}{\tB_k}&\qw\poloFantasmaCn{\rL}&\ghost{\tD_m}&\qw\poloFantasmaCn{\rM}&\multigate{1}{\tF_p}&\qw\poloFantasmaCn{\rN}&\ghost{\tG_q}\\
\pureghost{\Psi_i}&\qw\poloFantasmaCn{\rO}&\ghost{\tB_k}&\qw\poloFantasmaCn{\rP}&\qw&\qw
&\ghost{\tF_p}}\end{matrix}\quad
\begin{matrix}=\end{matrix}
$$
$$
\begin{matrix}\Qcircuit @C=1em @R=.7em @! R {
\multiprepareC{3}{\Psi_i}&\qw\poloFantasmaCn{\rA}&\multigate{1}{\tA_j}&\qw\poloFantasmaCn{\rB}&\qw&\\
\pureghost{\Psi_i}&\qw\poloFantasmaCn{\rE}&\ghost{\tA_j}&\qw\poloFantasmaCn{\rF}&\multigate{1}{\tD_m}&\qw\poloFantasmaCn{\rG}&\qw&\\
\pureghost{\Psi_i}&\qw\poloFantasmaCn{\rH}&\multigate{1}{\tB_k}&\qw\poloFantasmaCn{\rL}&\ghost{\tD_m}&\qw\poloFantasmaCn{\rM}&\multigate{1}{\tF_p}&\qw\poloFantasmaCn\rN&\qw&\\
\pureghost{\Psi_i}&\qw\poloFantasmaCn{\rO}&\ghost{\tB_k}&\qw\poloFantasmaCn{\rP}&\qw&\qw &\ghost{\tF_p}\\
}\end{matrix}
+
\begin{matrix}
\Qcircuit @C=1em @R=.7em @! R {&\qw\poloFantasmaCn{\rB} &\gate{\tC_l}&
\qw\poloFantasmaCn{\rC} &\multigate{1}{\tE_n}&\qw\poloFantasmaCn{\rD}&\multimeasureD{2}{\tG_q}\\
&&&\qw\poloFantasmaCn{\rG} &\ghost{\tE_n}&&\pureghost{\tG_q}\\
&&&&&\qw\poloFantasmaCn{\rN}&\ghost{\tG_q}}
\end{matrix}
$$
\caption{Split of a closed circuit into a preparation and an observation test.}\label{f:prepobs}
\end{figure}

\paragraph{Slices, preparation and observations.} Two wires in a circuit are input-output {\bf
  contiguous} if they are the input and the output of the same box. By following input-output
contiguous wires in a circuit while crossing boxes only in an input-output direction we draw an {\bf
  input-output chain}.  Two systems (wires) that are not in the same input-output path are called
{\bf independent}. A set of pairwise independent systems/wires is a {\bf slice}, and the slice is
called global if it partitions a closed bounded circuit into two parts as in Fig.
\ref{f:prepobs}
which, using our composition rules, is equivalent to the following
\begin{equation}
\Qcircuit @C=1em @R=.7em @! R {
\prepareC{(\Psi_i,\tA_j,\tB_k,\tD_m,\tF_p)}&\qw&\qw\poloFantasmaCn{\rB\rG\rN}&\qw
&\measureD{(\tC_l,\tE_n,\rG_q)}}
\end{equation}
namely, it ie equivalent to the connection of a preparation test with an observation test. Thus, a
diagram of the form $\Qcircuit @C=1em @R=.7em @! R {
  \prepareC{\tA_i}&\qw&\qw\poloFantasmaCn{\rA}&\qw &\measureD{\tB_j}}$ generally represents the
event corresponding to an istance of a concluded experiment, which starts with a preparation and
ends with an observation. The probability of such event will be denoted as $\SC{\tB_j}{\tA_i}$,
using the ``Dirak-like'' notation, with {\em rounded ket} $\K{\tA_i}$ and {\em rounded bra}
$\B{\tB_j}$ for the preparation and the observation tests, respectively.  In the following we will
use lowercase greek letters for preparation events and lowercase latin letters for observation
events. The following equivalent notations denote the probability of the sequence of events $\rho$,
$\tA$, $a$
\begin{equation}\label{lineareve1}
\B{a}\tA\K{\rho}=
\Qcircuit @C=1em @R=.7em @! R {
\prepareC{\rho} &\gate{\tA}&\measureD{a}}.
\end{equation}
Also,
\begin{equation}\label{lineareve2}
\Qcircuit @C=1em @R=.7em @! R {&\gate{\tA}&\measureD{a}}=
\Qcircuit @C=1em @R=.7em @! R {&\measureD{a\circ\tA}},\qquad \B{a}\tA=\B{a\circ\tA},
\end{equation}
and the event $\tA$ can be regarded as ``transforming'' the observation
event $a$ into the event $a\circ\tA$. The same can be said for the preparation event.

\subsection{States, effects, transformations}

In a probabilistic theory, a preparation-event $\rho_i$ for system $\rA$ is
naturally identified with a function sending observation-events of $\rA$ to probabilities, namely
\begin{equation}\label{statfun}
\rho_i: \Stset(\rA)\to [0,1], \quad \B{a_j}\mapsto \SC{a_j}{\rho_i},
\end{equation}
and, analogously, observation-events are identified with functions from pre\-paration-events to
probabilities
\begin{equation}
a_j: \Cntset(\rA)\to [0,1], \quad \K{\rho_i}\mapsto \SC{a_j}{\rho_i}.  
\end{equation}    
As probability rule, two observation-events (preparation-events) corresponding to the same function
are indistinguishable. We are thus lead to the following notions of states and effects:
\paragraph{States and effects:} {\em Equivalence classes of indistinguishable preparation-events for
  system $\rA$ are called {\bf states} of $\rA$, and their set is denoted as
  $\Stset(\rA)$. Equivalence classes of indistinguishable observation-events for system $\rA$ are
  called {\bf effects} of  $\sA$, and their set is denoted as $\Cntset(\rA)$.}

Therefore, in the following we will make the identifications: 1) preparation-events $\equiv$ states;
2) observation-events $\equiv$ effects. Notice that according to our definition of states and
effects as equivalence classes, states are separating for effects and viceversa effects are
separating for states.\footnote{We say that a set of effects is {\bf separating} for a set of
  states, if any two states of the set have at least a different probability for two effects of the
  other set. Similarly for a set of states.}

\paragraph{Linear spaces of states/effects.} Since states (effects) are functions from effects
(states) to probabilities, one can take linear combinations of them. This defines the real vector
spaces $\Stset_\Reals (\rA)$ and $\Cntset_\Reals (\rA)$, one dual of the other (we will restrict our
attention to finite dimensions).  In this case, by duality one has $\dim (\Stset_\Reals (\rA)) =
\dim (\Cntset_\Reals (\rA))$. 

\paragraph{Convex cones of states effects.} Linear combinations with positive coefficients of states
or of effects define the two convex cones $\Stset_{+} (\rA)$ and $\Cntset_{+} (\rA)$, respectively,
one dual cone of the other. The standard assumption in the literature is that, since the
experimenter is free to randomize the choice of devices with arbitrary probabilities, the set of
states $\Stset (\rA)$ and the set of effects $\Cntset (\rA)$ are convex. 

\paragraph{Linear extension of events.}
Linearity is naturally transferred to any kind of event through Eqs. (\ref{lineareve1}) and
(\ref{lineareve2}), via linearity of probabilities, and, in addition, events become linear maps on
states or effects, \eg $\tA\in\Trnset(\tA,\tB)$, $\tA:\K{\rho}_\rA\mapsto\K{\tA\rho}_\rB$.  Every
event $\tA\in\Trnset(\rA,\rB)$ induces a map from $\Stset(\rA\rC)$ to $\Stset(\rB\rC)$ for every
system $\rC$, uniquely defined by
\begin{equation}\label{extensA} 
\tA :\K{\rho}_{\rA \rC} \in \Stset (\rA \rC) \mapsto (\tA\otimes \tI_\rC) \K{ \rho}_{\rA \rC} \in
\Stset (\rB \rC),  
\end{equation}  
$\tI_\rC$ denoting the identity transformation on system $\rC$. The map is linear from
$\Stset_\Reals(\rA\rC)$ to $\Stset_\Reals (\rB\rC)$.  From a probabilistic point of view, if for
every possible system $\rC$ two events $\tA$ and $\tA'$ induce the same maps, then they are
indistinguishable. We are thus lead to the definition of {\bf transformation}: {\em Equivalence
  classes of indistinguishable events from $\rA$ to $\rB$ are called {\bf transformations} from
  $\rA$ to $\rB$.} Henceforth, we will identify events with transformations. Accordingly, a test
will be a collection of transformations.

\medskip In the following, if there is no ambiguity, we will drop the system index to the identity
event. Notice that generally two transformations $\tA,\tA' \in \Trnset (\rA , \rB)$ can be different
even if $\tA\K{ \rho}_\rA = \tA' \K{ \rho}_\rA$ for every $\rho \in \Stset (\rA)$. Indeed one has
$\tA\neq\tA'$ if that there exists an ancillary system $\rC$ and a joint state $\K{\rho}_{\rA \rC}$
such that
\begin{equation}
(\tA \otimes \tI)\K{\rho}_{\rA \rC} \not = (\tA'\otimes\tI)\K{ \rho}_{\rA \rC}.
\end{equation}
We will come back on this point when discussing local discriminability.

\subsection{No signaling from the ``future''.}
Although in the networks discussed until now we had sequences of tests, such sequences were not
necessarily temporal, or \emph{causal sequences}, namely the order of tests in a sequence was not
necessarily following the causal or the time arrow. 

We now introduce the causality condition, also called \emph{no signalling from the future}, which
allows us to interpret the sequential composition as a causal cascade.

\paragraph{Causality condition 1.}\cite{DAriano:2008p3785} {\em We say that a theory is {\em
    causal}, if for any two tests $\{\tA_i\}_{i\in\rX}$ and $\{\tB_j\}_{\j\in\rY}$ that are
  connected with at least an input of test $\{\tB_j\}_{\j\in\rY}$ connected to an output of
  $\{\tA_i\}_{\j\in\rY}$ as follows
\begin{equation}
\Qcircuit @C=1em @R=.7em @! R {&\ldots &&\ldots & &\qw
  \poloFantasmaCn \rC & \multigate{1}{\{\tB_j\}} & \qw \poloFantasmaCn \rF&\qw&\ldots\\ 
&\ldots && \qw \poloFantasmaCn \rA & \multigate{1} {\{\tA_i\}} & \qw
  \poloFantasmaCn \rD & \ghost {\{\tB_j\}} & \qw \poloFantasmaCn \rG&\qw&\ldots \\  
&\ldots && \qw \poloFantasmaCn \rB & \ghost {\{\tA_i\}} & \qw \poloFantasmaCn \rE &\qw& \ldots
&&\ldots} 
\end{equation} 
one has the asymmetry of the joint probability of events (given all other events in the network):
\begin{alignat}{2}\label{e:NSF}
&\sum_{j\in\rY}p(\tA_i,\tB_j)=p(\tA_i),\quad&&\forall\tA_i,\forall\{\tB_j\}_{j\in\rY},\\
&\sum_{i\in\rX}p(\tA_i,\tB_j)=p(\tB_j,\{\tA_i\}_{i\in\rX}),\quad&&\forall\{\tA_i\}_{i\in\rX},
\forall\tB_j.
\end{alignat}
In words, we say that the marginal over test $\{\tB_j\}_{j\in\rY}$ is independent on the choice of
the same test---namely it would be the same if there were no test at the output of test
$\{\tA_i\}_{j\in\rX}$---whereas the marginal over the test $\{\tA_i\}_{i\in\rX}$ generally depends
on the choice of the test.}

\paragraph{Causality and the arrow of time.} The above asymmetry of marginalization of joint
probabilities corresponds to say that: {\em test $\{\tA_i\}_{i\in\rX}$ can influence test
  $\{\tB_j\}_{j\in\rY}$, but not viceversa}. Or else: {\em $\{\tA_i\}_{i\in\rX}$ is cause for
  $\{\tB_j\}_{j\in\rY}$ and $\{\tB_j\}_{i\in\rY}$ is effect for $\{\tA_i\}_{i\in\rX}$}.  Thus, the
asymmetry is {\bf causality}. If we now take the input-output direction as the past-future time
relation, this corresponds to choose the {\bf arrow of time}, namely it corresponds to say that
causes precede effects.  According to our choice of the time-arrow the input-output connection
between tests is interpreted as a {\bf time-cascade of tests}. Therefore, in synthesis, the
asymmetry in the marginalization of probabilities corresponds to postulate that:

\medskip
{\bf No signaling from the future: }{\em The marginal probability of a time-cascade of tests does
  not depend on the ``future'' tests.} 

\medskip On the contrary, the marginal probability of a time-cascade of tests generally depends on
``past'' tests, and we will see that this leads to the customary probability-conditioning from the
past.

The causality condition greatly simplifies the evaluation of probabilities of events. In fact, since
the probability of an event in a test is independent on the tests performed at the output, we can
just substitute the network with another one in which all output systems of the test of interest are
substituted by a deterministic test.

\paragraph{Formulation in terms of preparation tests.} We have already introduced preparation tests,
namely tests with no input, and denoted as $\Qcircuit @C=1em @R=.7em @! R {\prepareC{\rho_i}& \qw
  \poloFantasmaCn \rB &\qw}$. Moreover, we have shown that every portion of network that has no
input is equivalent to a preparation test, as \eg in Fig. \ref{f:prepobs}. The causal condition can
now be equivalently formulated as follows:

\paragraph{Causal Condition 2.} \cite{Chiribella:2009unp} {\em A theory is \emph{causal} if every
  preparation-event $\K{\rho_j}_\rA$ has a probability $p(\rho_j)$ that is independent on the choice
  of test following the preparation test. Precisely, if $\{\tA_i\}_{i \in \rX}$ is an arbitrary test
  from $\rA$ to $\rB$, one has
\begin{equation}\label{no-sig}
p(\rho_j) = \sum_{i\in\rX}p(\tA_i\rho_j).
\end{equation}   
} The equivalence of the two formulations of the causal condition can be easily proved as follows.
The implication Condition 1 $\Longrightarrow$ Condition 2 is immediate.  Viceversa, consider any
portion of the complete network which has no input, which contains test $\{\tA_i\}_{i\in\rX}$, and
which has noting attached at the output systems of test $\{\tA_i\}_{i\in\rX}$, as follows
\begin{equation}
\Qcircuit @C=1em @R=.7em @! R {
\multiprepareC{3}{~^{}}&\multigate{1}{~^{}}&\gate{~^{}}&\qw\\
\pureghost{~^{}}&\ghost{~^{}}&\qw\poloFantasmaCn\rA&\multigate{1}{\{\tA_i\}}&\qw\poloFantasmaCn\rC&\qw\\
\pureghost{~^{}}&\multigate{1}{~^{}}&\qw\poloFantasmaCn\rB&\ghost{\{\tA_i\}}& \qw\poloFantasmaCn\rD&\qw\\
\pureghost{~^{}}&\ghost{~^{}}&\qw &}
\end{equation}
This is a preparation test. Then according to Condition 2 the joint probability of all events in the
preparation test---\ie our portion of network---is independent on the choice of tests connected at
the output of the network. In particular, the probability of events $\tA_i$ given all other events
in the network will be independent on the choice of the test at the output of test $\{\tA_i\}$.

\medskip
We should emphasize that there exist indeed input-output relation that have no causal
interpretation. Such non causal theories are studied in Ref. \cite{upcomingPaolo}. A concrete
example of such theories is that considered in Refs.  \cite{supermaps,comblong}, where the states
are quantum operations, and the transformations are ``supermaps'' transforming quantum operations
into quantum operations. In this case, transforming a state means inserting the quantum operation in
a larger circuit, and the sequence of two transformation is not a causal. The possibility of
formulating more general probabilistic theories even in the absence of a pre-defined causal arrow
may constitute a crucial ingredient for conceiving a quantum theory of gravity (see e.g. Hardy in
Ref.  \cite{Hardy:2007}).

The causality principle naturally leads to the notion of conditioned tests, generalizing both
notions of sequential composition and randomization of tests. For a precise definition see
Ref. \cite{Chiribella:2009unp}. 

Causal theories have a simple characterization in terms of the following lemma \cite{Chiribella:2009unp}.
\begin{lemma}\label{lem:charcaus} A theory is causal if and only if for every system $\rA$ there is
  a unique deterministic effect $\B{e}_\rA$. 
\end{lemma}
Moreover, one also has \cite{Chiribella:2009unp}
\begin{lemma}
  A theory where every state is proportional to a deterministic one is
  causal.
\end{lemma}

\subsection{Alternative definition of {\em state} for causal theories.} 
From Lemma \ref{lem:charcaus} it clear that in a causal theory the probability function over events
$p$ is uniquely defined. We can accordingly define the state also in the following way: {\em A {\bf
    state} $\omega$ for a system $\rA$ is a probability rule $\omega(\tA)$ for any event
  $\tA\in\Trnset(\rA,\rB)$ occurring in any possible test with input system $\rA$. We call the state
  {\bf normalized} if for every possible each test $\{\tA_i\}_{i\in\rX}$ with input system $\rA$,
  the following condition holds
\begin{equation}\label{nomega}
\sum_{j\in\rX}\omega(\tA_j)=1.
\end{equation}}

It is easy to see that for causal theories the above definition is equivalent to the definition of
state as equivalence class of preparation event. In fact, the preparation event is a positive
functional over observation tests (see Eq. (\ref{statfun})). On the other hand, due to causality,
the probability of the event $\tA$ for preparation $\K{\omega}_\rA$ is independent on the choice of
the following test, whence, in particular, is given by
\begin{equation}
\omega(\tA)={}_\rB\B{e}\tA\K{\omega}_\rA,
\end{equation}
whereas normalization follows easily. Viceversa, for a normalized state the probability rule
$\omega(\tA)$ along with normalization (\ref{nomega}) provides probabilities that satisfy
Eq. (\ref{no-sig}). 

\paragraph{Conditional state.} Causality also allows us to define the notion of conditional state,
namely the state corresponding to the conditional probability rule. The following cascade
\begin{equation}
\Qcircuit @C=1em @R=.7em @! R
{\prepareC{\omega}&\qw\poloFantasmaCn{\rA}&\gate{\tA}&\qw\poloFantasmaCn{\rB} &
\gate{\tB}&\qw\poloFantasmaCn{\rC}&\qw }
\end{equation}
leads to the notion of conditional probability that event $\tB$ occurs knowing that event $\tA$ has
occurred $p(\tB|\tA)=\omega(\tB\circ\tA)/\omega(\tA)$. This sets the new probability rule
$\omega_\tA(\tB):=p(\tB|\tA)$, corresponding to the notion of {\bf conditional state}: {\em The
  conditional state $\omega_\tA$, which gives the probability that an event occurs knowing that event
  $\tA$ has occurred with the system prepared in the state $\omega$, is given by
\begin{equation}
\omega_\tA\doteq\frac{\omega(\cdot\circ\tA)}{\omega(\tA)}.
\end{equation}}
(the central dot ``$\cdot$'' denotes the location of the variable). This is another way of regarding
the event $\tA$ as a transformation, namely as transforming with probability $\omega(\tA)$ the state
$\omega$ to the (unnormalized) state $\tA\omega$ given by
\begin{equation}
\tA\omega:=\omega(\cdot\circ\tA).
\end{equation}
In such way causality leads to the identifications: 1) event $\equiv$ transformation and 2)
evolution $\equiv$ state-conditioning. Notice that also a deterministic event produces a nontrivial
conditioning of probabilities. \bigskip

\paragraph{Marginal state.} Regarding the state as a probability rule in causal theories naturally
leads to the other relevant notion of {\bf marginal state}, corresponding to the marginalization
probability rule. The marginal state is just the probability rule for marginal probability, namely
\begin{equation}
\begin{split}
p_{ij}=&\B{a_i}_\rA\B{b_j}_\rB\K{\sigma}_{\rA\rB},\\
p_i=&\sum_jp_{ij}=\sum_j\B{a_i}_\rA\B{b_j}_\rB\K{\sigma}_{\rA\rB}=
\B{a_i}_\rA\B{e}_\rB\K{\sigma}_{\rA\rB}=:\B{a_i}_\rA\K{\rho}_\rA.
\end{split}
\end{equation}
Here $\K{\rho}_\rA$ is the marginal state of system $\rA$ of the joint state $\K{\sigma}_{\rA\rB}$. 
The definition of marginal state is therefore the following
\paragraph{Marginal state:} {\em The marginal state of  $\K\sigma_{\rA \rB}$ on system $\rA$ is the
  state $\K \rho_\rA := \B e_\rB \K\sigma_{\rA \rB}$.}  

\medskip
The marginalization of a state corresponds to the following diagram
\begin{equation}
\begin{matrix}\Qcircuit @C=1em @R=.7em @! R {
\multiprepareC{1}{\sigma}&\qw\poloFantasmaCn{\rA}&\qw\\
\pureghost{\sigma}&\qw\poloFantasmaCn{\rB}&\measureD{e}}\end{matrix}
\quad\begin{matrix}=:\end{matrix}\quad
\begin{matrix}
\Qcircuit @C=1em @R=.7em @! R {\prepareC{\rho}&\qw\poloFantasmaCn{\rA}&\qw}
\end{matrix}\,.
\end{equation}

\paragraph{Abbreviated notation.} In the following, when considering a transformation in
$\tA\in\Trnset(\rA,\rB)$ acting on a joint state $\omega\in\Stset(\rA\rC)$, we will think the
transformation acting on $\omega$ locally, namely we will use the following natural abbreviations
\begin{alignat}{2}
&\tA\in\Trnset(\rA,\rB),\;&\omega\in\Stset(\rA\rC),\qquad 
&\tA\omega\equiv(\tA\otimes\tI)\K{\omega}_{\rA\rC},\\ 
&&&\omega(\tA)\equiv \B{e}_{\rA\rC}(\tA\otimes\tI)\K{\omega}_{\rA\rC}. 
\end{alignat}
In regards of probabilities the abbreviation corresponds to take the marginal state.

\paragraph{Complete operational specification of a transformation.}
Operationally a transformation/event $\tA\in\Trnset(\rA,\rB)$ needs to be completely specified by
the way it affects all observed outcomes, namely all probabilities. This means that it is specified
by all the joint probabilities in which it is involved. It follows that $\tA\in\Trnset(\rA,\rB)$ is
univocally given by the probability rule 
\begin{equation}
\tA\in\Trnset(\rA,\rB),\quad\tA\omega=\omega(\cdot\circ\tA),\quad\forall \omega\in\Stset(\rA\rC)
\end{equation}
namely its local action on all joint states for any ancillary extensions.  This is equivalent to
specify both the conditional state $\omega_\tA$ and the probability $\omega(\tA)$ for all possible
states $\omega$, due to the identity
\begin{equation}\label{completespec}
\tA\omega=\omega(\tA)\omega_\tA.
\end{equation}
In particular the {\bf identity transformation} $\tI$ is completely specified by the rule
$\tI\omega=\omega$ for all states $\omega$. 

\paragraph{Linear space of events.}
We have seen that states inherite a linear structure from being functionals over effects, and the
viceversa effects inherite a linear structure from being functionals over states. We can also regard
the linear combination of two states as reflecting the linear combination of their respective
probability rule. On the other hand, since transformations/events are fully specified by their
action on states, they are also completely specified by their action over their linear space, hence
they inherit their linear structure as follows
\begin{equation}
(a\tA +b\tB)\omega:=a\tA\omega+b\tB \omega,\qquad \forall
a,b\in\Reals,\;\forall\omega\in\Stset(\rA\rC) 
\end{equation}
namely the linear combination of events $a\tA +b\tB$ is complely specified by its action over a
generic state $\omega\in\Stset(\rA\rC)$, action that is given by the linear combination of the two
states $\tA\omega,\tB \omega\in\Stset(\rA\rC)$. Notice that both compositions $\circ$ and $\otimes$
are distributive with respect to addition.

\subsection{No signaling without exchanging systems.}

The ``no signalling from the future'', \ie the causality requirement, implies another ``no
signaling'', namely the impossibility of signalling without exchanging systems:

\begin{theorem}{\bf (No signalling without exchange of physical
    systems)} In a causal theory it is impossible to have signalling
  without exchanging systems.
\end{theorem} 
\Proof  See Ref. \cite{Chiribella:2009unp}.

\subsection{Alternative definition of {\em effect} for causal theories.}

{\em An effect is the equivalence class of transformations occurring with the
  same probability}.

\medskip Indeed, if the two transformations $\tA_1,\tA_2\in\Trnset(\rA,\rB)$ are probabilistically
equivalent, one has
$\B{e}_\rA\tA_1\K{\omega}_\rA=\B{e}_\rA\tA_2\K{\omega}_\rA,\quad\forall\omega\in\Stset(\rA)$, and
due to the fact that states are separating for effects, this is equivalent to the identity of
effects $\B{e}_\rA\tA_1=\B{e}_\rA\tA_2:=\B{a}$, and we will say that the two transformations belong
to the same effect $a\in\Cntset(\rA)$.

\medskip\par Depending on the context, in the following we will also use the equivalent notations for
states, effects, and transformations
\begin{equation}
b\circ\tA=\B{b}\tA,\quad \tA\omega=\tA\K{\omega},\quad \B{b}\tA\K{\omega}=\omega(b\circ\tA).
\end{equation}

\medskip One of the consequences of Lemma \ref{lem:charcaus} is that the set of effects $\{l_i\}$
corresponding to all possible events of a test satisfy the normalization identity $\sum_il_i=e$, $e$
denoting the deterministic effect. Such a set of effects will be called {\bf observable}. We will
also call an observable {\bf informationally complete} if it is a state-separating set of effects,
and {\bf minimal}, if the effects are linearly independent.

\subsection{Local discriminability}

A standard assumption in the literature on probabilistic theories is \emph{local discriminability}. 

\paragraph{Local discriminability:} {\em 
A theory satisfies local discriminability if for every couple of different states $\rho, \sigma \in
\Stset (\rA \rB)$ there are two local effects $a \in \Cntset (\rA)$ and $b \in \Cntset (\rB)$ such
that}
\begin{equation}
\begin{matrix}\Qcircuit @C=1em @R=.7em @! R {\multiprepareC{1}{\rho}& \qw \poloFantasmaCn \rA
    &\measureD a \\  \pureghost\rho & \qw \poloFantasmaCn \rB &\measureD b}\end{matrix}
\;\begin{matrix}\not =\end{matrix}\;
\begin{matrix}\Qcircuit @C=1em @R=.7em @! R {\multiprepareC{1}{\sigma}& \qw
      \poloFantasmaCn \rA &\measureD a \\  \pureghost\sigma & \qw \poloFantasmaCn \rB &\measureD b}
\end{matrix}
\end{equation}

Another way of stating local discriminability is to say that the set of factorized effects is
separating for the joint states.

Local discriminability represents a dramatic experimental advantage. Without local discriminability,
one generally would need to built up a $N$-system test in order to discriminate an $N$-partite joint
state, instead of using just $N$ of the same single-system tests that allow us to discriminate
states of single system. Local discriminability implies local observability, namely the possibility
of recovering the full joint state from just local observations. Stated in other words, local
observability means that one can build up an informationally complete observation test made only of
local test, \ie one can perform a complete tomography of a multipartite state using only local
tests. This is given by the following lemma \cite{Chiribella:2009unp}:
\begin{lemma}
  Let $\{\rho_i\}$ and $\{\tilde \rho_j\}$ be two bases for the vector spaces $\Stset_\Reals (\rA)$
  and $\Stset_\Reals (\rB)$, respectively, and let $\{a_i\}$ and $\{b_j\}$ be two bases for the
  vector spaces $\Cntset_\Reals (\rA)$ and $\Cntset_\Reals (\rB)$, respectively. Then every state
  $\sigma \in \Stset (\rA \rB)$ and every effect $E \in \Cntset (\rA\rB)$ can be written as follows
\begin{equation}\label{linearcomb}
\begin{split}
\K{\sigma}_{\rA \rB} &= \sum_{i,j} ~A_{ij} \K{\rho_i}_{\rA} \K{\tilde \rho_j}_{\rB} \\
\B{E}_{\rA \rB} &= \sum_{i,j} ~B_{ij} \B{a_i}_{\rA} \B{b_j}_{\rB}
\end{split}
\end{equation}
for some suitable real matrix $A_{ij}$ ($B_{ij}$).
\end{lemma}  

Another consequence of local discriminability is that transformations in $\Trnset(\rA,\rB)$ are
completely specified by their action only on local states $\Stset(\rA)$, without the need of
considering ancillary extension. This is assessed by the following lemma \cite{Chiribella:2009unp}:
\begin{lemma}
If two transformations $\tA, \tB \in \Trnset (\rA, \rB)$ are different and local discriminability holds, then there exist a state $\rho\in \Stset (\rA)$ such that
\begin{equation}
 \Qcircuit @C=1em @R=.7em @! R { \prepareC \rho & \qw\poloFantasmaCn \rA & \gate \tA &\qw} ~\not =~  \Qcircuit @C=1em @R=.7em @! R { \prepareC \rho & \qw\poloFantasmaCn \rA & \gate \tB &\qw}
\end{equation}
\end{lemma}  

\section{Bloch representation for transformations of a probabilistic theory}\label{s:Blochrep}
Based on the linear structure established for states, effects, and transformations, we can now
introduce an affine-space representation based on the existence of a minimal informationally
complete observable and of a separating set of states. Such representation generalizes the popular
Bloch representation used in Quantum Mechanics.

\par In terms of a minimal informationally complete observable, $\{l_i\}$, $i=1,\ldots,n$,
and of a minimal separating set of states $\{\lambda_j\}$, $j=1,\ldots,n$, one can expand
(in a unique way) any effect $a\in\Cntset$ and state $\omega\in\Stset$ as follows
\begin{equation}
a=\sum_{j=1}^N\lambda_j(a)l_j,\qquad\omega=\sum_{j=1}^Nl_j(\omega)\lambda_j.
\end{equation}
Instead of using a minimal informationally complete observable and a minimal set of separating
states it is convenient to adopt canonical biorthogonal basis $\mat l=\{l_i\}$ and
$\mat\lambda=\{\lambda_j\}$ for $\Cntset_\Reals$ and $\Stset_\Reals$ embedded into $\Reals^n$ as
Euclidean space, and it is convenient to identify an element in $\{l_i\}$ with the deterministic
effect $e$---say $l_n=e$. Correspondingly $\lambda_n$ in $\mat\lambda=\{\lambda_j\}$ is the
functional $\chi$ giving the deterministic component of the effect. Using a Minkowskian notation
\begin{equation}
  \mat l\doteq(\mat{\hat l},e),\;\mat\lambda\doteq(\mat{\hat\lambda},\chi)
  \qquad\text{with}\qquad\lambda\cdot l\doteq \sum_j\lambda_jl_j=\mat{\hat\lambda}\cdot\mat{\hat l}+\chi e,
\end{equation}
we write
\begin{equation}\label{e:Mink_scal}
(a,\omega)=\omega(a)=a(\omega)=\mat l(\omega)\cdot\mat\lambda(a):=\sum_{i=1}^nl_i(\omega)\lambda_i(a)
\equiv\mat{\hat\lambda}(a)\cdot\mat{\hat l}(\omega)+\chi(a)e(\omega).
\end{equation}
Clearly one can extend the convex sets of effects and states to their complexification by taking complex expansion coefficients.

The vectors $\mat l(\omega)$ and $\mat\lambda(a)$ give a complete description of the (unnormalized)
state $\omega$ and (unbounded) effect $a$, thanks to identity (\ref{e:Mink_scal}). For normalized
state $\omega$, $\mat l(\omega)$ is the \emph{Bloch vector} representing the state
$\omega$\footnote{More precisely the last component of $\mat l(\omega)$ is $e(\omega)=1$ for each
  normalized $\omega$, and the Bloch vector is $\mat{\hat l}(\omega)$.}. The representation is {\em
  faithful} (\ie one-to-one) for biorthogonal basis or, generally, for minimal informationally
complete observable.
\par We now recover the linear transformation describing conditioning. The conditioning is given by
$\B{b}\tA\K{\omega}=
\tA\omega(b)=\omega(b\circ\tA)= b(\tA\omega)$. From linearity of transformations one can introduce a matrix $\mat A\equiv\{A_{ij}\}$, and write 
\begin{equation}\label{e:matrix_repr}
l_i(\tA\omega)=\omega(l_i\circ\tA)=\sum_{j=1,\ldots,n-1} A_{ij}l_j(\omega)+ A_{in}e(\omega),
\end{equation}
and, in particular, upon denoting $a:=[\tA]_\eff$, one has
\begin{equation}
e(\tA\omega)=\omega(e\circ\tA)\equiv\omega(a)=\sum_j A_{nj}l_j(\omega)
\equiv \mat{\hat\lambda}(a)\cdot\mat{\hat l}(\omega)+\chi(a)e(\omega),
\end{equation}
from which we derive the identities $\lambda_j(a)\equiv A_{nj}$ and $\chi(a)=A_{nn}$.

\begin{figure}[ht]\label{fig:matA}
$$
\mat A=\begin{pmatrix}
\fbox{\parbox[t][22mm][c]{22mm}{\begin{center}$\mat{\hat A}$\end{center}}}
&\!\!\!\!\fbox{\parbox[t][22mm][c]{12mm}{\begin{center}$\mat{\hat k}(\tA)$\end{center}}}\\
\fbox{\parbox[t][12mm][c]{22mm}{\begin{center}$\transp{\mat{\hat \lambda}(a)}$\end{center}}} 
& \!\!\!\!\fbox{\parbox[t][12mm][c]{12mm}{\begin{center}$\chi(a)$\end{center}}}
\end{pmatrix},\quad
\begin{matrix}
\mat{\hat l}(\tA\omega)=\mat{\mat  A}\mat{\hat l}(\omega)+\mat{\hat k}(\tA),\\ \\
\omega(\tA)=\mat{\hat\lambda}(a)\cdot\mat{\hat l}(\omega)+\chi(a),\\ \\
\displaystyle{\mat{\hat l}(\omega_\tA)=\frac{\mat{\hat A}(\tA)\mat{\hat l}(\omega)+\mat{\hat k}(\tA)}{\mat{\hat\lambda}(a)\cdot\mat{\hat l}(\omega)+\chi(a)}}.
\end{matrix}
$$
\caption[Matrix representation of the real algebra of transformations $\aA$.]{Matrix
  representation of the real algebra of transformations $\aA$. The last row represents the effect
  $a$ of the transformation $\tA$. It gives the transformation of the zero-component of the Bloch
  vector $e(\tA\omega)\equiv\omega(\tA)=\mat{\hat\lambda}(a)\cdot\mat{\hat l}(\omega)+\chi(a)$, namely the
  probability of the transformation. The other rows represent the affine transformation of the
  Bloch vector $\mat{\hat l}(\omega)$ corresponding to the operation of $\tA$ , the last column
  giving the translation $\mat{\hat k}(\tA)$, and the remaining square matrix $\mat{\hat A}$ the linear
  part of the affine map. The Bloch vector of the state $\omega$ is transformed as $\mat {\hat l}(\tA\omega)=\mat{\hat A}\mat{\hat l}(\omega)+\mat{\hat k}(\tA)$, and the conditioning over the convex set of states is the fractional affine transformation in figure.}
\end{figure}

The real matrices $\mat A$ are a \emph{representation} of the real algebra of generalized transformations $\aA$.  The last row of the matrix is a representation of the effect $a=[\tA]_{\text{eff}}$ (see Fig. \ref{fig:matA}). In vector notation, for a normalized input state one has
\begin{equation}\label{e:transfBloch}
\begin{split}
\mat l(\tA\omega)=&\mat{\hat A}\mat l(\omega)+\mat{\hat k}(\tA),\qquad
\mat{\hat k}(\tA)\doteq\mat{\hat l}(\tA\chi)\\
e(\tA\omega)=&\mat{\hat\lambda}(a)\cdot\mat{\hat l}(\omega)+\chi(a),\\
\tA\omega(b)=&\mat{\hat\lambda}(b)\cdot\mat{\hat l}(\tA\omega)+\chi(b)e(\tA\omega).
\end{split}
\end{equation}
The matrix representation of the transformation is given in Fig. \ref{fig:matA}.

\par Therefore, summarizing, we have found the following representation for the conditional state $\omega_\tA$ after the action of the transformation $\tA$ regarded as an affine map over $\Stset$ 
\begin{equation}
\omega\in\Stset,\quad\mat{\hat l}(\omega)\longrightarrow\mat{\hat l}(\omega_\tA)=\frac{\mat{\hat A}\mat{\hat l}(\omega)+\mat{\hat k}(\tA)}{\mat{\hat\lambda}(a)\cdot\mat{\hat l}(\omega)+\chi(a)},
\end{equation}
with the transformation occurring with probability $\omega(\tA)$ given by
$\omega(\tA)=\mat{\hat\lambda}(a)\cdot\mat{\hat l}(\omega)+\chi(a)$.
Naturally is
\begin{equation}
\omega\in\Stset,\quad\mat l(\omega)\longrightarrow\mat l(\omega_\tA)=\Bigg(\frac{\mat{\hat A}\mat{\hat l}(\omega)+\mat{\hat k}(\tA)}{\mat{\hat\lambda}(a)\cdot\mat{\hat l}(\omega)+\chi(a)},1\Bigg).
\end{equation} 
A pictorially view of the action over $\Stset$ of the affine map $\tA$ is given by the \emph{linear-fractional map} and the \emph{perspective map} (see \cite{Boyd:ConvOpt}).

\par The following Propositions will be useful in constructing concrete probabilistic models.

\begin{proposition}\label{prop:Bloch_repr1}
All the contractions in the convex set $\Trnset$ are represented in Bloch form by a matrix with an element of $\Cntset$ as last row. Otherwise they will not be contractions.
\end{proposition}
\Proof By definition of Bloch representation.\qed

\begin{proposition}\label{prop:Bloch_repr2}
If $a\in\Extr(\Cntset)$ and $\tA\in\Extr(a)$ then $\tA\in\Extr(\Trnset)$.
\end{proposition}
\Proof If $\tA\in a$ then its Bloch matrix has $\mat\lambda(a)$ as last row. According to
Proposition \ref{prop:Bloch_repr1} every set of contractions combining convexly to give $\tA$ must
combine to $\mat\lambda(a)$ in the last row of the Bloch representation. Since $a\in\Extr(\Cntset)$,
the only case in which it is possible is when the convex combination is among elements in the same
equivalence class $a$, but this contradicts the hypothesis $\tA\in\Extr(a)$.\qed

\begin{observation}\label{obs:1}
  One could think that all extremal transformations are extremal within the equivalence class $a$
  with $a$ extremal, namely $\Extr(\Trnset)=\{\tA\in\Extr(a),\forall a\in\Extr(\Cntset)\}$. In
  general this is false, as we will show with an example from Quantum Mechanics. On the contrary, we
  will see that the extended Popescu-Rohrlich model satisfies this property.
\end{observation} 
\Definition{We define the {\bf generator set of $\Cntset$}---denoted by $\set{gen}(\Cntset)$---as the set of effects whose orbit under the group of $\Cntset$-automorphisms is $\Cntset$, namely the set such that $\set{gen}(\Cntset)\circ\set{Aut}(\Stset)=\Cntset$.}
\begin{proposition}\label{prop:Bloch_repr3}
We get $\Extr(a\circ\set{Aut}(\Cntset))=\Extr(a)\circ\set{Aut}(\Cntset)$ $\forall\, a\in\set{gen}(\Cntset)$.
\end{proposition}
\Proof It is sufficient to show that $\forall\, a\in\set{gen}(\Cntset)$ and $\forall\tU\in\set{Aut}(\Stset)$ we get
\begin{equation}\label{e:extr_aut}
\Extr(a\circ\tU)=\Extr(a)\circ\tU.
\end{equation}
Considering a map $\tB$ in $\Extr(b)$ it is easy to show that $\tB\circ\tU$ is a map in $\Extr(b\circ\tU)$, in fact the Bloch representative of $\tB\circ\tU$ has $\mat\lambda(b\circ\tU)$ as last row and is extremal because $\tB=\tB\circ\tU\circ\tU^{-1}$ is extremal. Then for each $\tA\in\Extr(a\circ\tU)$ in Eq. (\ref{e:extr_aut}) we can take $\tA\circ\tU^{-1}\in\Extr(a)$ satisfying the equality, and vice-versa for each $\tA\in\Extr(a)$ we can take $\tA\circ\tU\in\Extr(a\circ\tU)$.\qed

In the following we will denote by $\mat l=\{l_i\}$ and $\mat\lambda=\{\lambda_j\}$ the canonical
basis of $\Cntset_\Reals$ and $\Stset_\Reals$, respectively.

\section{The Postulates FAITH, FAITHE, and PURIFY}

\subsection{Postulate PFAITH}\label{ss:PFAITH} Postulate PFAITH playes a major role in all
operational probabilistic theories (see both Refs. \cite{DAriano:2008p3785} and
\cite{Chiribella:2009unp}). The Postulate concerns the possibility of calibrating any test and of
preparing any joint bipartite state only by means of local transformations.  Before introducing the
Postulate we need to define what is a faithful state.

\par
Consider a bipartite system $\rA\rB$ and a bipartite state $\Phi\in\Stset(\rA\rB)$.  The state
$\Phi$ induces the following cone-homomorphism\footnote{A cone-homomorphism between the cones $K_1$
  and $K_2$ is simply a linear map between $\Span_\Reals(K_1)$ and $\Span_\Reals(K_2)$ which sends
  elements of $K_1$ to elements of $K_2$, but not necessarily vice-versa.}
\begin{equation}\label{e:cone_homo}
\Trnset_+(\rA)\ni\tA\mapsto(\tA\otimes\tI)\Phi\in\Stset_+(\rA\rB).
\end{equation}
\begin{itemize}
\item If the cone-homomorphism in Eq. (\ref{e:cone_homo}) is a cone-monomorphism, namely the output
  $(\tA\otimes\tI)\Phi$ is in one to one correspondence with the local transformation $\tA$, then $\Phi$
  is {\bf dinamically faithful} with respect to $\rA$. The output keeps the information about
  the input transformation and this allows to calibrate any test by means of local transformations.
\item If the cone-homomorphism in Eq. (\ref{e:cone_homo}) is a cone-epimorphism, namely every
  bipartite state $\Psi$ can be achieved as $\Psi=(\tA_\Psi\otimes\tI)\Phi$ for some local transformation
  $\tA_\Psi$, then $\Phi$ is {\bf preparationally faithful} with respect to $\rA$. Any joint
  state can be prepared by means of local transformations.
\end{itemize}

\begin{observation} For $\Phi$ both preparationally and dynamically faithful, one can operationally
  define the {\bf transposed transformation} $\tA^\prime\in\Trnset_\Reals(\rA)$ of a transformation
  $\tA\in\Trnset_\Reals(\rA)$ through the  identity
\begin{equation}
(\tA^\prime\otimes\tI)\Phi=(\tI\otimes\tA)\Phi,
\end{equation}
and all the properties of transposition are verified.
\end{observation}

\Postulate{PFAITH}{Existence of a symmetric preparationally faithful pure state}{ For any couple of
  identical systems, there exists a symmetric (under permutation of the two systems) bipartite state
  which is both pure and preparationally faithful} Postulate PFAITH leads to many relevant features
of the probabilistic theory. Here we briefly report those that are useful in the construction of our
concrete probabilistic models. For the proof see Ref.\cite{DAriano:2008p3785} where many other
consequences are investigated.  In the following, when considering two identical systems $\rA=\rB$
if there is no ambiguity we will just write $\rA\rA$ instead of $\rA\rB$ to denote the bipartite
system. Consider a probabilistic theory for two identical systems $\rA=\rB$ that satisfies Postulate
PFAITH and let $\Phi$ be a pure symmetric and preparationally faithful bipartite state of the
theory; then the following properties holds:
\begin{itemize}
\item[(1)] $\Phi$ is both preparationally and dinamically faithful with respect to both systems.
\item[(2)]One has the cone-isomorphism\footnote{Two cones $K_1$ and $K_2$ are isomorphic iff there
    exists a linear bijective map between the linear spans $\Span_\Reals(K_1)$ and
    $\Span_\Reals(K_2)$ that is cone preserving in both directions, namely it and its inverse map
    must send $\Erays(K_1)$ to $\Erays(K_2)$ and positive linear combinations to positive linear
    combinations.} $\Trnset_+(\rA)\simeq\Stset_+(\rA\rA)$ induced by $\Phi$ via the map
  $\tA\in\Trnset_+(\rA)\leftrightarrow(\tA\otimes\tI)\Phi\in\Stset_+(\rA\rA)$. Moreover, a local
  transformation on $\Phi$ produces an output pure (unnormalized) bipartite state iff the
  transformation is atomic, namely $\Psi=(\tA_\Psi\otimes\tI)\Phi$ is pure iff $\tA_\Psi$ is atomic.
\item[(3)]The theory is weakly self-dual, namely one has the cone-isomorphism
  $\Cntset_+(\rA)\simeq\Stset_+(\rA)$ induced by the map $\Phi(a,\cdot)=\omega_a$ $\forall
  a\in\Cntset_+(\rA)$.
\item[(4)]The identical transformation $\tI$ is atomic.
\item[(5)]The transpose of a physical automorphism of the set of states is still a physical automorphism of the set of states. We denote the set of automorphism of the set of states by $\set{Aut}(\Stset(\rA))$.
\item[(6)]The maximally chaotic state $\chi\coloneqq\Phi(e,\cdot)$ is invariant under the transpose
  of a channel (deterministic transformation) whence, in particular, under a physical automorphism
  of the set of states.
\end{itemize}
\begin{observation} A stronger version of PFAITH, satisfied by Quantum Mechanics, requires the
  existence of a symmetric preparationally {\bf superfaithful} state $\Phi$, such that also
  $\Phi\otimes\Phi$ is preparationally faithful, whence $\Phi^{\otimes 2n}$ is
  preparationally faithful with respect to $\rA^n$, $\forall n>1$. 
\end{observation}

\subsection{Postulate FAITHE and teleportation}
In Ref.\cite{DAriano:2008p3785} other Postulates are introduced which make the probabilistic
theories closer to Quantum Mechanics. In this paper these Postulates will be tested on concrete
probabilistic models.
\par Since a preparationally faithful state is also dynamically faithful, it is ideed an
isomorphism, and as a matrix it is invertible. On the other hand, in general its inverse is not a
bipartite effect: \Postulate{FAITHE}{Existence of a faithful effect}{ There exists a bipartite
  effect $F$ (all system equal) achieving probabilistically the inverse of the cone-isomorphism
  $\Cntset_+(\rA)\simeq\Stset_+(\rA)$ given by $a\rightarrow\omega_a\coloneqq\Phi(a,\cdot)$, namely
\begin{equation}\label{e:FAITHE}
\B{F}_{23}\K{\omega_a}_2=\B{F}_{23}\B{a}_1\K{\Phi}_{12}=\alpha \B{a}_3,\quad 0\leq\alpha\leq 1.
\end{equation}
} Eq. (\ref{e:FAITHE}) is equivalent to $\B{F}_{23}\K{\Phi}_{12}=\alpha\tS_{13}$, $\tS_{ij}$
denoting the transformation which swaps the $i$th system with the $j$th system. The main consequence
of FAITHE is the possibility of achieving {\bf probabilistic teleportation} of states between equal
systems using the effect $F$ and the state $\Phi$ as follows
\begin{equation}
\begin{split}
\B{F}_{23}\K{\omega}_2\K{\Phi}_{34}=&
\B{F}_{23}\B{a_\omega}_1\K{\Phi}_{12}\K{\Phi}_{34}=
\B{a_\omega}_1\underbrace{\B{F}_{23}\K{\Phi}_{12}}_{\tS_{13}}\K{\Phi}_{34}\\=&
\alpha\K{a_\omega}_1\K{\Phi}_{14}=\alpha\K{\omega}_4.
\end{split}
\end{equation}
According to the last equation Postulate FAITHE is equivalent to the relation
\begin{equation}\label{e:FAITHE_equiv}
\B{F}_{23}\K{\Phi}_{12}\K{\Phi}_{34}=\alpha\K{\Phi}_{14},\quad\text{or}\quad
\end{equation}
where $\alpha$ is the probability of achieving teleportation. It depends only on the faithful effect
$F$ since it is $\alpha=\B{e}_{14}\B{F}_{23}\K{\Phi}_{12}\K{\Phi}_{34}$. Moreover, the maximum value of $\alpha$ is
achieved maximizing over all bipartite effects and it depends on the particular probabilistic
theory.
\par Here we give a criterion to exclude the possibility of achieving teleportation from a
preparationally faithful state in a probabilistic theory.
\begin{proposition}\label{prop:FAITH_viol}If there exists a preparationally faithful state violating
  Postulate FAITHE then all the preparationally faithful states violate it.
\end{proposition}
\Proof Let $\Phi\in\Stset(\rA\rA)$ be the preparationally faithful state violating
FAITHE. And let $F=\alpha\Phi^{-1}$ be the bipartite functional satisfying Eq. (\ref{e:FAITHE}).
Then there exists a state $\Psi\in\Stset(\rA\rA)$ such that
\begin{equation}
\SC{\Phi^{-1}}{\Psi}<0,
\end{equation}
namely $F=\alpha\Phi^{-1}$ is not a true effect for each $\alpha>0$.  Now let $\Phi^\prime$ be
another preparationally faithful state. From the faithfulness of $\Phi$, there exists a
transformation $\tA\in\Trnset(\rA)$ such that
\begin{equation}
(\tA\otimes\tI)\Phi=\Phi^\prime.
\end{equation}
$\Phi^\prime$ is preparationally and dinamically faithful, therefore $\tA$ is invertible and
$\tA^{-1}$ is a transformation. Consider then the quantity
\begin{equation}
\SC{\Phi^{-1}}{\Psi}=\B{\Phi^{-1}}(\tA\otimes\tI)(\tA^{-1}\otimes\tI)\K{\Psi}
=\SC{{\Phi^\prime}^{-1}}{\Psi^\prime}<0.
\end{equation}
So also ${\Phi^\prime}^{-1}$ is not a bipartite effect because we have found a state
$\Psi^\prime=(\tA^{-1}\otimes\tI)\Psi\in\Stset(\rA\rA)$ such that
$\SC{{\Phi^{-1}}{^\prime}}{\Psi^\prime}<0$.\qed \bigskip As immediate consequence of this Proposition we
get

\begin{corollary}\label{cor:FAITH_tel_cor}If a probabilistic theory does not satisfy Postulate FAITHE then there is no preparationally faithful state achieving teleportation.
\end{corollary}

\par The following Proposition will be useful in the construction of probabilistic models because it shows that a model which violates Postulate FAITHE cannot admit the existence of a super-faithful state.

\begin{proposition}\label{prop:super_faith}
  If a probabilistic theory admits a super-faithful state $\Phi$, then Postulate FAITHE is
  automatically satisfied and teleportation is achievable.
\end{proposition}
\Proof In fact considering the symmetric faithful quadripartite state $\Phi_{quad}=\Phi\otimes\Phi$,
according to the isomorphism $\Cntset_+(\rA\rA)\simeq\Stset_+(\rA\rA)$, we can find a
bipartite effect $F_\Phi\in\Cntset(\rA\rA)$ such that
\begin{equation}\label{e:sup_faith}
\B{F_\Phi}_{23}\K{\Phi}_{12}\K{\Phi}_{34}=\alpha\K{\Phi}_{14},
\end{equation}
as required by FAITHE (see Eq. (\ref{e:FAITHE_equiv})). Naturally teleportation follows as a
consequence of Postulate FAITHE.\qed

\subsection{Purifiability of a probabilistic theory} We know that Quantum Mechanics allows
purification. A ``minimal'' version of purifiability for probabilistic theories is introduced
through the following Posulate: 
  \Postulate{PURIFY}{Purifiability of all states}{For every state $\omega\in\Stset(\rA)$ there
    exists a purification $\Omega\in\Stset(\rA\rA)$, i.~e. namely a state $\Omega$ having $\omega$ as
    marginal state. Precisely:
\begin{equation}
  \forall\omega\in\Stset(\rA),\quad\exists\Omega\in\Extr(\Stset(\rA\rA)),\quad\text{such that}
\quad\B{e}_2\K{\Omega}_{12}=\K{\omega}_1.
\end{equation}}
In Ref.\cite{Chiribella:2009unp} many consequences of PURIFY are analysed. In particular is proved the following Lemma which achieves the atomicity of the identical transformation, and then the purity of the preparationally faithful state, without assuming PURIFY: 
\begin{lemma}\label{lem:PURIFY_Id_Atomicity} If Postulate PURIFY holds then the identical
  transformation is atomic and the preparationally faithful state $\Phi$ is pure.
\end{lemma}

\par As already mentioned Postulate PURIFY introduces a minimal notion of purifiability. Quantum Mechanics satisfies a more restrictive condition. Therefore in the same work Ref.\cite{Chiribella:2009unp} is introduced a {\bf stronger version of Postulate PURIFY} which requires that every mixed state has a purification, unique up to reversible channels on the purifying system. This new property has the \emph{entanglement swapping} (and then probabilistic teleportation) as a consequence:
\begin{proposition}\label{prop:entsw} Consider a probabilistic theory. If every mixed state has a purification, unique
  up to reversible channels on the purifying system, then each symmetric pure bipartite
  preparationally faithful state $\Phi\in\Stset(\rA\rA)$ allows entanglement swapping.  Thus FAITHE
  is satisfied and probabilistic teleportation is achievable.
\end{proposition}
For the proof see Ref.\cite{Chiribella:2009unp}. Notice that the stronger version of Postulate
PURIFY requires the uniqueness of purification up to reversible channels on the purifying system at
all the multipartite levels.  Given a faithful state $\Phi\in\Stset(\rA\rA)$ we say that
the {\bf entanglement swapping} is possible if there exists a constant $\alpha>0$ and a bipartite
effect $F\in\Cntset(\rA\rA)$ such that
\begin{equation}
\B{F}_{23}\K{\Phi}_{12}\K{\Phi}_{34}=\alpha\K{\Phi}_{14}.
\end{equation}
Therefore, according to Eq. (\ref{e:FAITHE_equiv}) FAITHE is satisfied and teleportation is achievable.

\bigskip
In the following sections we will test the above postulates on sime probabilistic toy-theories different
from Quantum Mechanics.

\section{Toy-theory 1: the two-box world (extended Popescu-Rohrlich model)}
The original model contains only states and effects, and has been already considered in
Ref.\cite{DAriano:2008p3785} as a testing model for our present probabilistic framework. Here we
will extend the model, by adding transformations in a consistent fashon.

\subsection{Original model: the Popescu-Rohrlich boxes}
\begin{figure}[ht]
 \begin{minipage}[b]{5cm}
   \centering
   \includegraphics[width=1.4cm]{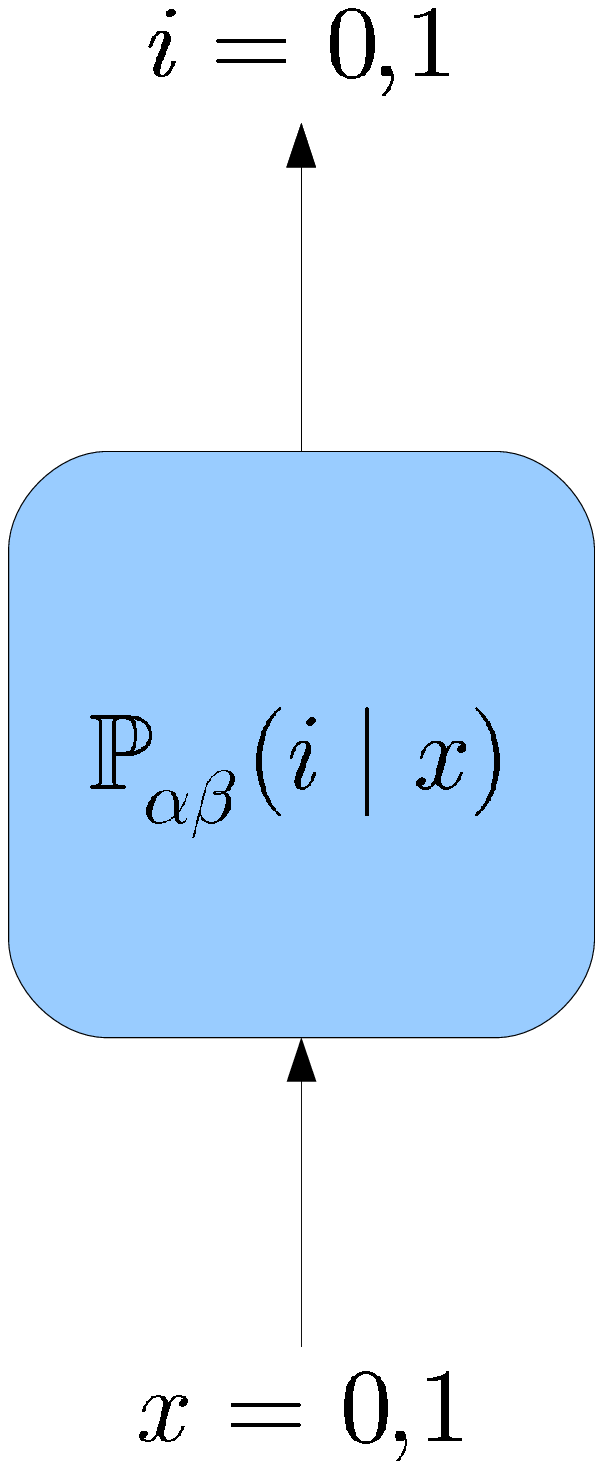}
\caption{}\label{fig:prb}
 \end{minipage}
 \begin{minipage}[b]{7cm}
  \centering
   \includegraphics[width=4cm]{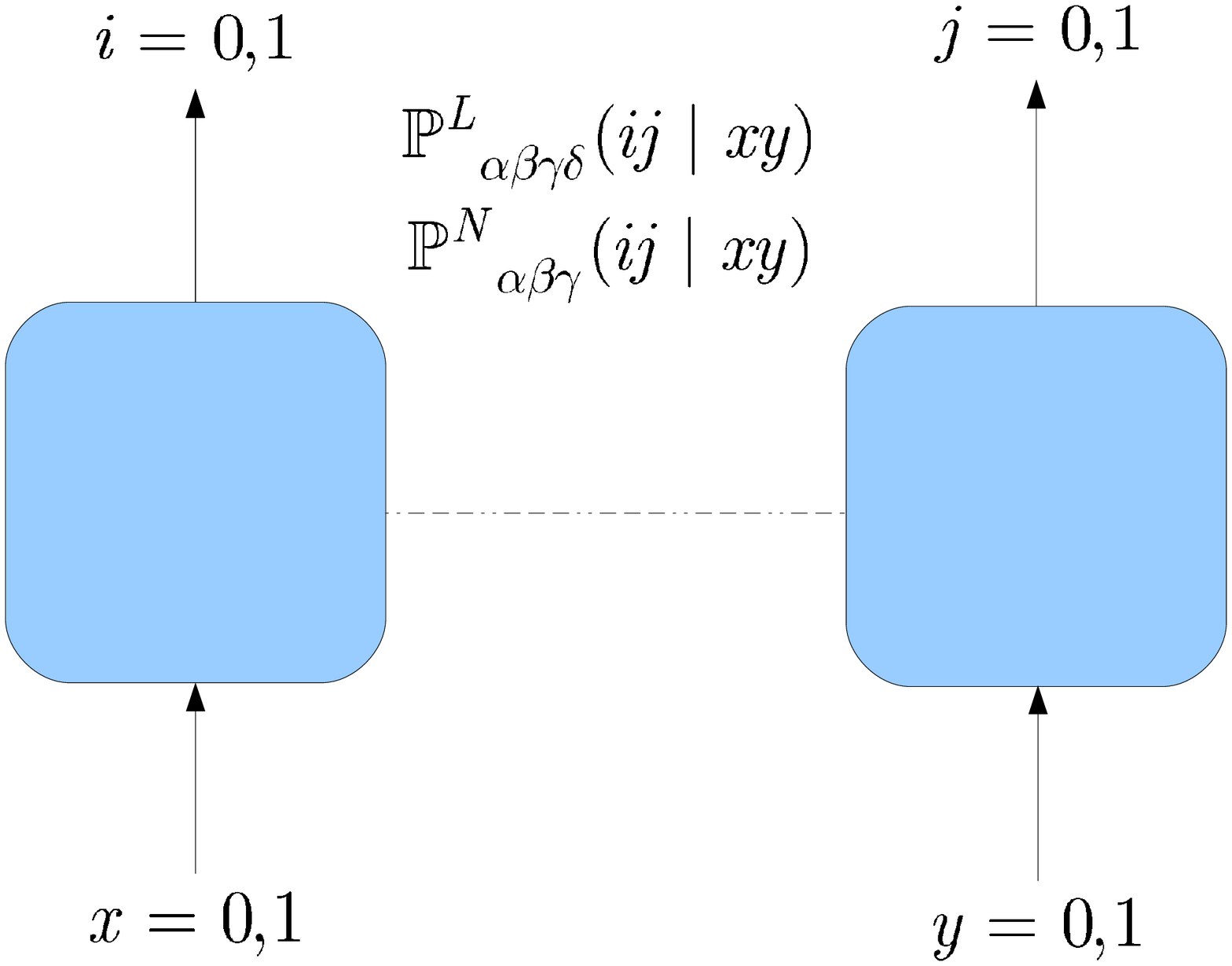}
\caption{}\label{fig:prbb}
 \end{minipage}
\end{figure}

The riginal model has ben presented in Ref.\cite{Rohrlich:1995p1940}.  It is locally made of
a {\emph box} which provides the probability rule for the output given the input. In the simplest
situation, input and output are both binary. As sketched in Fig. \ref{fig:prb}, the
probability rules are\footnote{In Eq. (\ref{e:PR}) the symbol $\oplus$ denotes the addition modulo $2$.}
\begin{equation}\label{e:PR}
\Prob_{\alpha\beta}(i|x)=\begin{cases}1, & i=\alpha x\oplus\beta\\
0&\text{otherwise},\end{cases}\qquad\alpha,\beta=0,1,
\end{equation}
for the two possible outputs $i=0,1$ given the two possible inputs $x=0,1$. 
\par The core of the original work are the correlated boxes in Fig. \ref{fig:prbb} defined by the
joint probabilities $\Prob(ij|xy)$ consistent with no-signaling. As shown in
Ref.\cite{Barrett:2004}, the complete set of such probabilities make an eight dimensional polytope
with 24 vertices. Among these 24 probability distributions we can identify the two relevant classes
of local non-local boxes, denoted as $\Prob_{\alpha\beta\gamma\delta}^{L}(ij|xy)$ and
$\Prob_{\alpha\beta\gamma\delta}^{N}(ij|xy)$, respectively:
{\small
\begin{equation}\label{e:PRj}
\Prob_{\alpha\beta\gamma\delta}^{L}(ij|xy)=\begin{cases}1,  & i=\alpha x\oplus\beta\\&j=\gamma x\oplus\delta\\ 
0&\text{otherwise},
\end{cases},\quad
\Prob_{\alpha\beta\gamma}^{N}(ij|xy)=\begin{cases}\frac{1}{2},  & i\oplus j=xy\oplus\alpha x\oplus\beta
  y\oplus\gamma\\ 
0&\text{otherwise},
\end{cases}
\end{equation}}
where $\alpha,\beta,\gamma,\delta\in\{0,1\}$. The 16 local vertices
$\Prob_{\alpha\beta\gamma\delta}^{L}(ij|xy)$ correspond to the factorization of the single box
probability rules $\Prob_{\alpha\beta}(i|x)$, while the 8 non-local probability rules
$\Prob_{\alpha\beta\gamma}^{N}(ij|xy)$ introduce the strongest correlations compatible with
no-signaling, corresponding to the maximal violation of the CHSH inequality with no-signaling.

\par In the following we will introduce the cones of states and effects, and then we will extend the original model introducing transformations. This will be achieved starting from a bipartite state considered as preparationally faithful.
 
\subsection{Local sets of states and effects}\label{ss:PR_loc_sets}
According to the local box in Fig. \ref{fig:prb} we can perform two possible tests, $\AA^{(x)}=\{\tA_0^{(x)},\tA_1^{(x)}\}$ with $x=0,1$, and, correspondingly, we will denote the effects of the test $\AA^{(x)}$ as $a_0^{(x)},a_1^{(x)}$, with 
\begin{equation}
a_0^{(0)}+a_1^{(0)}=a_0^{(1)}+a_1^{(1)}=e,
\end{equation}
where $e$ is the deterministic effect. Therefore there are only three independent local effects,
whence $\dim(\Cntset_+)=\dim(\Stset_+)=3$. Clearly $\dim(\Stset)=2$, and there are only two affinely
independent states. Therefore, the local convex set of states is the 2-dimensional \emph{polytope}
${\mathbb P}^2$ given by the convex hull of the probability rules $\Prob_{\alpha\beta}(i|x)$ in Eq.
(\ref{e:PR}). These are the vertices of $\Stset$, namely the pure states of the model. In the
following we will denote them by $\omega^{\alpha\beta}$.
   
\par It is convenient to represent the effects in a $3$-dimensional vector space with the canonical
coordinate along the $z$-axis corresponding to the deterministic effect $e$. Therefore a possible
representation of the four effects in the two tests is

\begin{equation}\label{e:square_eff}
\mat\lambda(e)=\begin{bmatrix}0\\ 0\\ 1\end{bmatrix},
\mat\lambda(a_0^{(0)})=\begin{bmatrix}\frac{1}{2}\\ -\frac{1}{2}\\ \frac{1}{2}\end{bmatrix},\,
\mat\lambda(a_1^{(0)})=\begin{bmatrix}-\frac{1}{2}\\\frac{1}{2}\\ \frac{1}{2}\end{bmatrix},\,
\mat\lambda(a_0^{(1)})=\begin{bmatrix}\frac{1}{2}\\ \frac{1}{2}\\ \frac{1}{2}\end{bmatrix},\,
\mat\lambda(a_1^{(1)})=\begin{bmatrix}-\frac{1}{2}\\-\frac{1}{2}\\ \frac{1}{2}\end{bmatrix}.
\end{equation}
Correspondingly, according to the probability rule in Eq. (\ref{e:PR}), the four pure states will be represented as
\begin{equation}
\mat l(\omega^{00}):=\begin{bmatrix}1\\ 0\\1 \end{bmatrix},\,\mat l(\omega^{11}):=\begin{bmatrix}0\\ 1\\ 
    1\end{bmatrix},\,\mat l(\omega^{01}):=
\begin{bmatrix}-1\\ 0\\1 \end{bmatrix},\,\mat l(\omega^{10}):=\begin{bmatrix}0\\ -1\\ 1\end{bmatrix}.
\end{equation}
One can easily verify the application of the states to the effects
\begin{equation}
\Prob_{\alpha\beta}(i|x)=\omega^{\alpha\beta}(a_i^{(x)})\equiv \mat l(\omega^{\alpha\beta})\cdot
  \mat\lambda(a_i^{(x)})=\begin{cases}1, & i=\alpha x\oplus\beta\\ 
    0&\text{otherwise}.\end{cases}
\end{equation}
\begin{figure}[ht]
\centering
\includegraphics[width=0.3\textwidth]{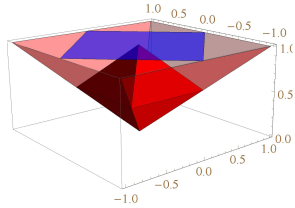}
\caption{The square at the top represents the set of states $\Stset$. The transparent cone represents the dual cone of effects $\Cntset_+$. The octahedron inside the transparent cone represents the convex set of effects $\Cntset$ which is the $\Cntset_+$-truncation given by the condition $a\leq e$ where $a$ is a generic effect and $e$ the deterministic one}
\label{fig:PR_cones}
\end{figure}
Notice the third coordinate (the axis of the cone $\Stset_+$), which is constantly equal to unit.
Denoting by $x,y,z$ the three components of vectors in both the Euclidean spaces $\Stset_\Reals$ and
$\Cntset_\Reals$, the $2$-dimensional polytope $\mathbb P^2$ of states is (see the square at the top
in Fig. \ref{fig:PR_cones})
\begin{equation}
\Stset={\mathbb P}^2=\Big\{\mat l(\omega)\;|\;|x|+|y|\leq 1\Big\},
\end{equation}
which is the convex hull of the vectors $\mat l(\omega^{\alpha\beta})$ ($\alpha=0,1,\,\beta=0,1)$ corresponding to the vertices of $\Stset$.  Clearly the cone $\Stset_+$, based on $\Stset$, and its dual $\Cntset_+$ are given by
\begin{equation}
\Stset_+=\Big\{\mat l(\omega)\,|\,|x|+|y|\leq z,\, z\geq 0\Big\},\quad
\Cntset_+=\Big\{\mat\lambda(a)\,|\,|x|\leq z,\, |y|\leq z,\,z\geq 0\Big\}.
\end{equation}
Therefore the convex set of physical effects is  
\begin{equation}
\Cntset=\left\{\mat\lambda(a)\:\,\text{such that}\:
\begin{cases}|x|\leq z,\,|y|\leq z,&z\in\big[0,\frac{1}{2}\big]\\
|x|\leq 1-z,\,|y|\leq 1-z,&z\in\big[\frac{1}{2},1\big]\end{cases}
\right\},
\end{equation}
corresponding to the truncation of $\Cntset_+$ given by the order prescription $0\leq a\leq e$. 

\subsection{The bipartite system and the faithful state}
As mentioned, the joint probabilities $\Prob_{\alpha\beta\gamma}(ij|xy)$ form a table of $2^{4}=16$
entries, of which only 8 of them are independent. Thus the bipartite convex set of states
$\Stset(\rA\rA)$ is the 8-dimensional polytope with the 24 vertices given in Eq. (\ref{e:PRj}).
These correspond to the pure bipartite states of the model: the 16 factorized states
$\omega^{\alpha\beta}\otimes\omega^{\gamma\delta}$, plus the 8 non-local ones, which we will denote
by $\Phi^{\alpha\beta\gamma}$.  The whole set $\Stset(\rA\rA)$ is then the convex hull of its
vertices. A way to introduce the whole set of transformations\footnote{We know that transformations
  are usually the completely positive maps but in this model as in the following ones we consider
  only two systems and then the transformations are two-positive maps by construction.} $\Trnset_+$
compatible with the cone of bipartite states $\Stset_{+}(\rA\rA)$, is to assume the
cone-isomorphism $\Trnset_+(\rA)\simeq\Stset_{+}(\rA\rA)$ induced by a preparationally faithful state
$\Phi$ according to Postulate PFAITH. We can take one of the non-local vertices
$\Phi^{\alpha\beta\gamma}$---say $\Phi=\Phi^{000}$---as a pure symmetric preparationally faithful
state. First we have to check that, regarded as a matrix over effects, such state is non singular,
since a preparationally faithful state is also an isomorphic map between the cones $\Stset_+$ and
$\Cntset_+$. Indeed we have
\begin{equation}
\Phi=\sum_{ij}\Phi_{ij}\lambda_i\otimes\lambda_j\equiv\mat{\Phi}=
\{\Phi_{ij}\}=\{\Phi(l_i,l_j)\},
\end{equation}
and from the rules in the right Eq. (\ref{e:PRj}) we get the non singular matrix
\begin{equation}\label{e:matr_repr}
\mat\Phi=\footnotesize\begin{bmatrix}\Phi(a_0^{(0)}-a_1^{(1)},a_0^{(0)}-a_1^{(1)}) & \Phi(a_0^{(0)}-a_1^{(1)},a_1^{(0)}-a_0^{(0)}) & \Phi(a_0^{(0)}-a_1^{(1)},e)\\
{}&{}&{}\\
\Phi(a_0^{(0)}-a_1^{(1)},a_1^{(0)}-a_0^{(0)}) & \Phi(a_1^{(0)}-a_0^{(0)},a_1^{(0)}-a_0^{(0)}) & \Phi(a_1^{(0)}-a_0^{(0)},e)\\
{}&{}&{}\\
\Phi(a_0^{(0)}-a_1^{(1)},e) & \Phi(a_1^{(0)}-a_0^{(0)},e) & \Phi(e,e)
\end{bmatrix}=\tfrac{1}{2}\begin{bmatrix} 1 & -1 & 0\\ -1 & -1 & 0\\ 0 & 0 & 2 \end{bmatrix}\normalsize.
\end{equation}
The cone-isomorphism $\Cntset_+\simeq\Stset_+$ established by the map $\Phi(a,\cdot)=\omega_a$ is explicitly given by $\varphi_i:=\Phi(l_i,\cdot)$, where the vectors $\varphi_i$ are the images of the basis effects $l_i$  under the map $\Phi$. One also has
\begin{equation}
\Phi(a_0^{(0)},\cdot)=\tfrac{1}{2}\omega^{00},\quad
\Phi(a_1^{(0)},\cdot)=\tfrac{1}{2}\omega^{01},\quad
\Phi(a_0^{(1)},\cdot)=\tfrac{1}{2}\omega^{10},\quad
\Phi(a_1^{(1)},\cdot)=\tfrac{1}{2}\omega^{11}.
\end{equation}
Notice that $\Phi(e,\cdot)=\chi$ has representative $\mat l(\chi)=\lambda_3$, namely it is the center of the square $\Stset$. 
\par The same arguments leading to the matrix representation of $\Phi^{000}$ can be iterated for
each state $\Phi^{\alpha\beta\gamma}$, and all of them could be assumed as faithful state of the
theory.

\subsection{Introducing transformations}\label{ss:transf} As already stated the symmetric preparationally faithful state $\Phi^{000}$ induces the cone-isomorphism $\Trnset_+(\rA)\simeq\Stset_{+}(\rA\rA)$. The first step is to achieve from the isomorphism an explicit relation between elements in the two cones. Then by this relation the whole set $\Trnset_+$ could be generated from the cone of bipartite states $\Stset_{+}(\rA\rA)$.
Let $\tA$ be a generic transformation in $\Trnset_+$. Then take the matrix representation of $\tA$ induced by the relation
\begin{equation}
l_i\circ\tA=:\sum_kA_{ik}l_k.
\end{equation}
From the isomorphism $\Trnset_+\simeq\Stset_{+}(\rA\rA)$ we know that
\begin{equation}
\forall\;\Psi\in\Stset_+(\rA\rA)\qquad\exists!\;\tA_\Psi\in\Trnset_+\;\text{ such that }\;(\tI\otimes\tA_\Psi)\Phi=\Psi.
\end{equation}
Matching the last two equations we have
\begin{equation}\label{e:expl_isom}
\begin{split}
(\tI\otimes\tA_\Psi)\Phi(l_i,l_j)=\Psi(l_i,l_j)\;&\Rightarrow\;
\Phi(l_i,l_j\circ\tA_\Psi)=\Psi(l_i,l_j)\;\Rightarrow\;
\sum_k\Phi_{ik}A_{ik}=\Psi_{ij}\\
&\Rightarrow\mat\Phi\transp{\mat A}_\Psi=\mat\Psi\Rightarrow\mat A_\Psi=\transp{\mat\Psi}\mat\Phi^{-1}.
\end{split}
\end{equation}
It is sufficient to find the twenty-four extremal rays of $\Trnset_+$, namely the ones associated to the extremal rays of $\Stset_{+}(\rA\rA)$, according to the cone-isomorphism $\Trnset_+(\rA)\simeq\Stset_{+}(\rA\rA)$.

First we achieve the transformations corresponding to the non-local vertices $\Phi^{\alpha\beta\gamma}$, namely the eight maps $\tD^{\alpha\beta\gamma}$ such that
\begin{equation}
(\tI\otimes\tD^{\alpha\beta\gamma})\Phi=\Phi^{\alpha\beta\gamma}
\qquad\alpha,\beta,\gamma=0,1.
\end{equation}
From their representatives $\mat D^{\alpha\beta\gamma}=\Phi^{\alpha\beta\gamma}\Phi^{-1}$ it is easy to verify the identity $\set {Aut}(\Stset)=\{\tD^{\alpha\beta\gamma}\}$, namely the maps $\tD^{\alpha\beta\gamma}$ are the eight automorphisms of the local square of states $\Stset$: $\tD^{000},\tD^{111},\tD^{001},\tD^{110}$ perform respectively a $2\pi$, $\pi/2$, $\pi$, $3\pi/2$, $2\pi$-clockwise rotations around the axis of the cone $\Stset_+$, while $\tD^{100},\tD^{011},\tD^{010},\tD^{101}$ perform the four reflection-symmetries of the square of states.  
As a consequence of PFAITH (see Subsec. \ref{ss:PFAITH}) the transposed of the automorphisms must be still automorphisms as can be directly verified in this case. Moreover the application of the automorphisms to the faithful state $\Phi$ produces the eight pure bipartite states of $\Stset(\rA\rA)$ which are all pure symmetric preparationally faithful states. Finally we can verify that the maximally chaotic state $\chi=\Phi_{000}(e,\cdot)$ is invariant under the automorphisms application, namely $\tD^{\alpha\beta\gamma}\chi=\chi$, $\forall\;\tD^{\alpha\beta\gamma}\in\set{Aut}(\Stset)$, as stated among the PFAITH consequences in Subsec. \ref{ss:PFAITH}.
\par The other extremal elements of $\Trnset_+$ are the transformations associated to the sixteen pure states $\omega^{\alpha\beta}\otimes\omega^{\gamma\delta}$. From the explicit isomorphism in Eq. (\ref{e:expl_isom}) we get sixteen transformations, the eight maps  
\begin{equation}\label{e:extr_tr}\footnotesize
\tfrac{1}{2}\b{c}{-c}{c}{0}{0}{0}{1}{-1}{1},\quad
\tfrac{1}{2}\b{-c}{c}{c}{0}{0}{0}{-1}{1}{1},\quad
\tfrac{1}{2}\b{c}{c}{c}{0}{0}{0}{1}{1}{1},\quad
\tfrac{1}{2}\b{-c}{-c}{c}{0}{0}{0}{-1}{-1}{1},\qquad\normalsize
c=\pm 1,
\end{equation}
plus the eight given by inverting the first and the second rows.  
From these transformations plus the automorphisms $\{\tD^{\alpha\beta\gamma}\}$ it is possible to generate the extremal rays of the cone $\Trnset_+$ ($\Erays(\Trnset_+)$) and, by convex combinations, the whole set $\Trnset_+$.
\par As already mentioned in Observation \ref{obs:1}, the extended Popescu-Rohrlich model has the
following interesting property

\begin{proposition} The extremal transformations of the extended Popescu-Rohrlich model coincide
  with the extremal elements of the equivalences classes if extremal effects.
\end{proposition}
\Proof We know from Subsec. \ref{ss:PR_loc_sets} that
$\Extr(\Cntset)=\{e,0,a_0^{(0)},a_0^{(1)},a_1^{(0)}a_1^{(1)}\}$. According to the definition of
$\set{gen}(\Cntset)$ given in Sec. \ref{s:Blochrep}, we can assume
$\set{gen}(\Cntset)=\{e,0,a_0^{(0)}\}$. In fact acting on $a_0^{(0)}$ with the automorphisms
$\set{Aut}(\Stset)$ the remaining extremals of $\Cntset$ are achieved.
\par First we look for $\Extr(a_0^{(0)})$. From Proposition \ref{prop:Bloch_repr1} we know that the
Bloch representative $\mat A=\{A_{ij}\}$ of a transformation $\tA\in a_0^{(0)}$ has
$\mat\lambda(a_0^{(0)})$ as last row, namely it is $A_{31}=\tfrac{1}{2}$, $A_{32}=-\tfrac{1}{2}$ and
$A_{33}=\tfrac{1}{2}$.  Moreover $\tA$ must be positive and then
$\tA\omega^{\alpha\beta}\in\Stset_+$ $\forall\alpha\beta\in{0,1}$. Remembering the definitions of
$\omega^{\alpha\beta}$ and $\Stset_{+}$ the last conditions produce the four inequalities
\begin{equation}\label{e:bound}
\begin{split} 
&|A_{11}+A_{13}|+ |A_{21}+A_{23}|\leq 1,\quad
|-A_{12}+A_{13}|+|-A_{22}+A_{23}|\leq 1,\\
&|A_{12}+A_{13}|+|A_{22}+A_{23}|\leq 0,\quad
|-A_{11}+A_{13}|+ |-A_{21}+A_{23}|\leq 0.
\end{split}
\end{equation}
The third and the last bounds fix the equalities $A_{12}=-A_{13}$, $A_{21}=-A_{23}$, $A_{11}=A_{13}$
and $A_{21}=A_{23}$ making the positivity condition as simple as $|A_{11}|+
|A_{21}|\leq\frac{1}{2}$. The extremals of this set of matrixes, namely $\Extr(a_0^{(0)})$, are the
four maps
\begin{equation}\label{e:extr_a00}
\begin{bmatrix} c & c & c\\0 & 0 & 0\\\frac{1}{2} & -\frac{1}{2} & \frac{1}{2}\end{bmatrix},\qquad\begin{bmatrix} 0 & 0 & 0\\c & c & c\\\frac{1}{2} & -\frac{1}{2} & \frac{1}{2}\end{bmatrix}.\qquad c=\pm\frac{1}{2}.
\end{equation}
According to Proposition \ref{prop:Bloch_repr3} the extremals in the equivalence classes $a_0^{(1)}$, $a_1^{(0)}$ and $a_1^{(1)}$ follows from the application of $\set{Aut}(\Stset)$ to the matrixes in Eq. (\ref{e:extr_a00}). The result are exactly the sixteen maps associated to the sixteen pure states $\omega^{\alpha\beta}\otimes\omega^{\gamma\delta}$ by the cone isomorphism $\Stset_+^{\otimes 2}\simeq\Trnset_+$.
\par Finally we consider the Bloch representatives of the deterministic transformations in $e$, whose last row is $\mat\lambda(e)=[0,0,1]$. A simple calculus, similar to the previous one, shows that $\Extr(e)$ are exactly the eight automorphisms $\set{Aut}(\Stset)=\{\tD^{\alpha\beta\gamma}\}$ associated by the cone isomorphism $\Stset_+^{\otimes 2}\simeq\Trnset_+$ to the eight pure states $\Phi^{\alpha\beta\gamma}$.\qed

\subsection{Impossibility of teleportation} It is well known that the Popescu-Rohrlich model
exhibits stronger nonlocality than Quantum Mechanics. For this reason one may argue that
teleportation should be achievable. However, this is not the case, as we will see in the following.
Consider for example the preparationally faithful state $\Phi^{000}=\Phi$ and the bilinear form $F$
such that
\begin{equation}\label{e:tp}
\B{F}_{23}\K{\Phi}_{12}\K{\Phi}_{34}=\alpha\K{\Phi}_{14},
\end{equation}
for some $\alpha\in(0,1]$. In order to satisfy Eq.\ (\ref{e:tp}) the matrix $\mat F$, which
represents $F$ in our Bloch basis, must be proportional to $\mat \Phi^{-1}$, namely
\begin{equation}
F\propto(l_{1}\otimes l_{1})-(l_{1}\otimes l_{2})-(l_{2}\otimes l_{1})-(l_{2}\otimes l_{2})+(l_{3}\otimes l_{3}).
\end{equation}
It is easy to verify that $F$ is not a genuine bipartite effect. In fact, while the application of
$F$ to separable states always gives positive result
\begin{equation}
F(\omega,\zeta)\geq0 \qquad\forall\: \omega,\zeta\in\Stset,
\end{equation}
exploring the application of $F$ to bipartite states, we find
\begin{equation}
F(\Phi_{001})\propto\Phi_{001}(l_{1},l_{1})-\Phi_{001}(l_{1},l_{2})
-\Phi_{001}(l_{2},l_{1})-\Phi_{001}(l_{2},l_{2})+\Phi_{001}(l_{3},l_{3})=-1
\end{equation} 
This shows that Postulate FAITHE is not satisfied in this model and, according to Corollary
\ref{cor:FAITH_tel_cor}, teleportation cannot be achieved in the extended Popescu-Rohrlich
probabilistic theory.
\begin{observation} Notice that according to Proposition \ref{prop:super_faith} the Popescu-Rohrlich
  theory does not admit a super-faithful state, which, instead, would achieve probabilistic
  teleportation.
\end{observation}
  
\subsection{A theory without purification}\label{ss:PR_purification} 
Another fundamental quantum feature, the purifiability of all states, is not satisfied by the
Popescu-Rohrlich model, namely Postulate PURIFY does not hold. In fact the only pure bipartite
states, apart from the sixteen factorized ones $\omega^{\alpha\beta}\otimes\omega^{\gamma\delta}$,
are the eight maximally correlated states in Eq. (\ref{e:PRj}) which are all purifications of the
maximally chaotic state $\chi$
\begin{equation}
\Phi_{\alpha\beta\gamma}(\cdot,e)=\Phi_{\alpha\beta\gamma}(e,\cdot)=
\chi\qquad\forall\alpha,\beta,\gamma=0,1.
\end{equation}
In conclusion there are too few pure bipartite states with respect to the infinite mixed states to be purified (the internal points of the square $\Stset$). This will not be the case in the following class of models.
     
\section{Toy-theory 2: the two-clock world}
The Two-clock probabilistic models have a {\bf clock} as local system, namely a system with convex
set of states which is the disk $\mathbb B^{2}$. Many theories with such a local convex set of
states set can be generated: here we investigate their properties as probabilistic theories.

\subsection{The self-dual local system}\label{ss:local_TC_sd}
\begin{figure}[ht]
 \begin{minipage}[c]{5cm}
   \centering
   \includegraphics[width=4.5cm]{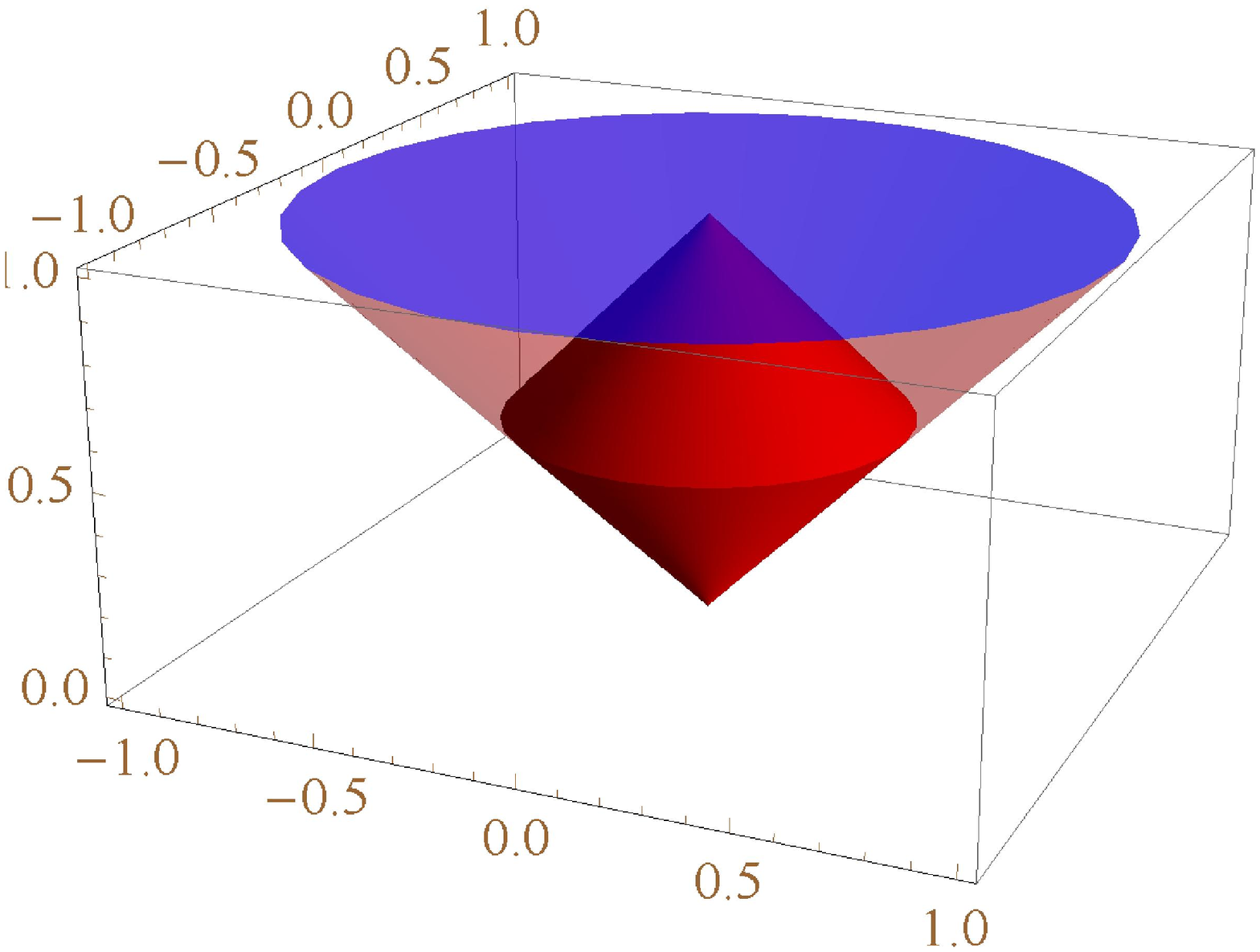}
 \end{minipage}
 \ \hspace{2mm} \hspace{2mm} \
 \begin{minipage}[c]{6.5cm}
  \centering
   \includegraphics[width=5.5cm]{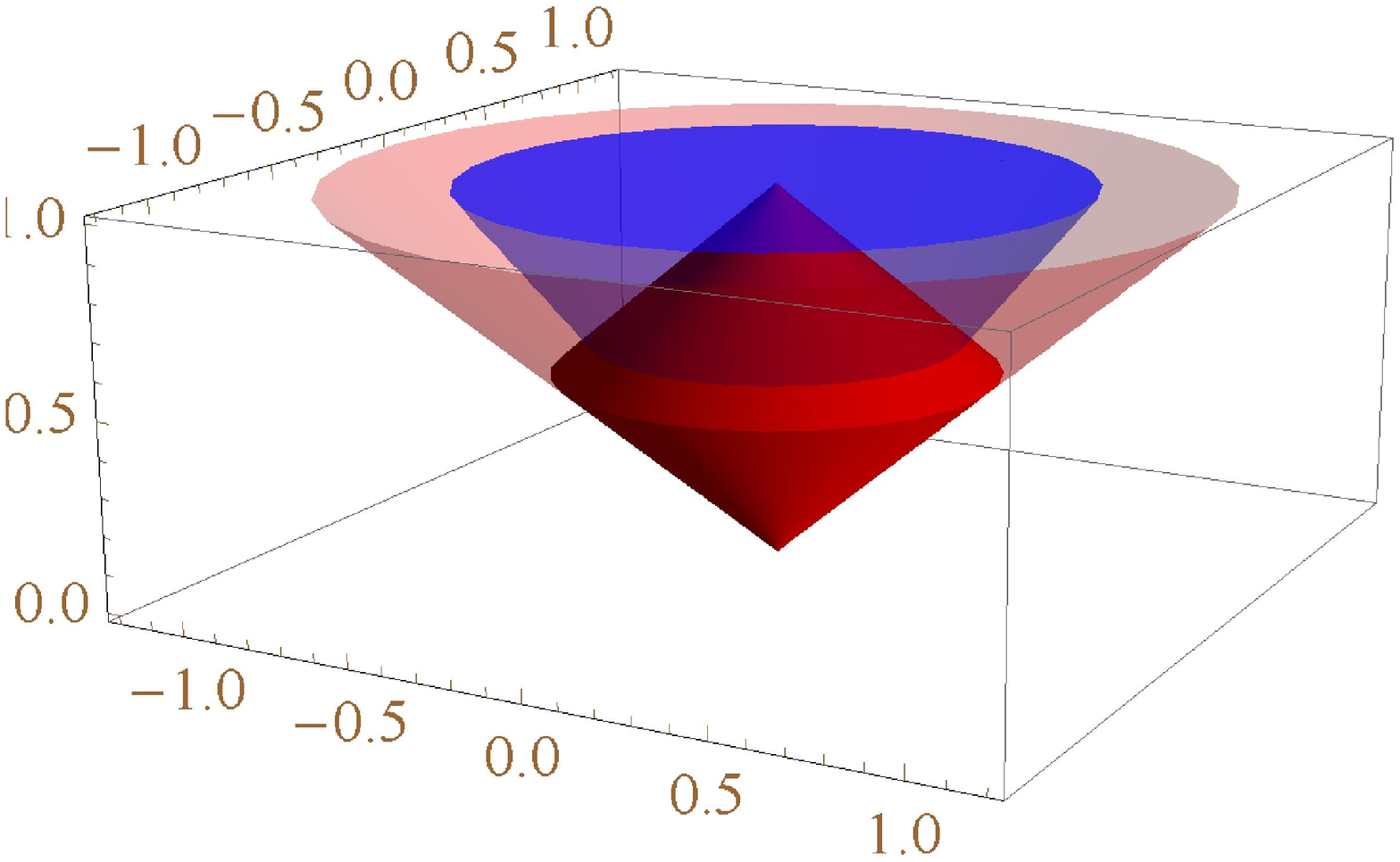}
 \end{minipage}
 \caption{{\bf Left figure:} the disk at the top represents the set of states $\Stset$. The
   transparent cone represents the cones $\Stset_+$ and $\Cntset_+$. The solid inside the
   transparent cone represents the convex set of effects $\Cntset$ which is the
   $\Cntset_+$-truncation given by the condition $a\leq e$ where a is a generic effect and e is the
   deterministic one. {\bf Right figure:} the same as in the left figure in the non self-dual
   case.}\label{fig:TC_cones_sd}
\end{figure}

We can consider the model self-dual at the local system level. Therefore, in the usual
representation, the cones of states and effects coincide
\begin{equation}\label{e:tcsc}
\Stset_+=\left\{\mat l(\omega)\;|\;x^2+y^2\leq z^2,\;z\geq 0\right\},\quad
\Cntset_+=\left\{\mat\lambda(a)\;|\;x^2+y^2\leq z^2,\;z\geq 0\right\}, 
\end{equation}
namely the theory is (pointedly) self-dual at a single system level if we embed both cones in the
same Euclidean space $\Reals^3$. As usual, the deterministic effect in our canonical basis is given
by the vector $\mat\lambda(e)=[0,0,1]$.  The set of states $\Stset\equiv{\mathbb B^2}$ is the basis
of the cone $\Stset_+$ at $z=1$, whereas the convex set of effects $\Cntset$ is the set of points of
the cone $\Cntset_+$ satisfying $e-a\in\Cntset_+$, namely
\begin{equation}\label{e:tcss}
\Stset=\left\{\mat l(\omega)\;|\;x^2+y^2=1\right\},\quad
\Cntset=\left\{\mat\lambda(a)\;|\;
x ^2+y^2\leq \min(z^2,(1-z)^2),\;z\in[0,1]\right\}.
\end{equation}
Therefore, the convex set of effects $\Cntset$ is made of two truncated cones of height $\frac{1}{2}$
glued together at the basis, as in the left Fig. \ref{fig:TC_cones_sd}, with the two vertices given by the null and the deterministic effect.

\subsection{The faithful state choice}\label{ss:faithst}
Differently from the two-box world, the model doesn't provide the joint states, which we will now
construct. Although the local cones do not identify uniquely the bipartite system, its structure
will be tightly connected to the local one, if the model has a faithful state. In fact a faithful
state must provide the automorphism $\Stset_+\simeq\Cntset_+$ between the local cones of states and
effects, thus narrowing the possible choices for the faithful state itself.  Let's introduce the
bipartite functional
\begin{equation}\label{e:TC_faith_state}
\Phi(a,b)=\sum_{i}\lambda_i(a)\lambda_i(b).
\end{equation}
One can check that it is positive over the cone of effects, but also over its linear span. $\Phi$ can be taken as a pure preparationally faithful state. Indeed, $\Phi$ gets the cone-isomorphism
$\Stset_+\simeq\Cntset_+$, via the map
\begin{equation}
\omega_a:=\Phi(a,\cdot)=a.
\end{equation}
in agreement with self-duality. Notice that, similarly to the Popescu-Rohrlich model, the
deterministic effect corresponds to the state $\chi=\Phi(e,\cdot)=\lambda_3$ at the center of
$\Stset$ .
\par In the two-box world we have generated $\Trnset_+$ from the given cone $\Stset_+(\rA\rA)$
using the isomorphism $\Stset_+(\rA\rA)\simeq\Trnset_+(\rA)$ induced by the preparationally faithful
state of the theory.  Here we choose the cone of physical transformations $\Trnset_+$ and use
isomorphism induced by $\Phi$ to deduce the cone of bipartite states $\Stset_+(\rA\rA)$. The
explicit isomorphism is that of Eq. (\ref{e:expl_isom}), namely $\mat
A_\Psi=\transp{\mat\Psi}\mat\Phi^{-1}$. Now each bipartite state has the same representative matrix
of the corresponding transposed transformation because, in terms of the canonical basis one has
$\Phi=\sum_{i=1}^3\lambda_i\otimes\lambda_i$, that is $\mat\Phi=I_3$. Thus the isomorphism simply
reads
\begin{equation}\label{e:expl_isom_TC}
\mat\Psi=\transp{\mat A}_\Psi.
\end{equation}

\subsection{Physical transformations}\label{ss:TC_transf}
We are left with the problem of searching among the positive maps, which are also two-positive:
these will be the physical maps of our model. The extremal transformations $\Erays(\Trnset_+)$ are
the maps sending $\Extr(\Stset)$ into an elliptical conic of $\Erays(\Stset_+)$,\footnote{A conic
  section (or just a conic) is a curve obtained by intersecting a cone (more precisely a circular
  conical surface) with a plane.}, which we will call \emph{elliptical-maps}. There exist three
different kinds of elliptical-maps corresponding to the three different elliptical conics:
\begin{enumerate}
    \item[{\bf a.}] \emph{Circular-maps.} In these case the map $\tA$ sends $\Extr(\Stset)$ into a circle (which is a particular ellipse) and then $\Stset$ into a disk. 
    \item[{\bf b.}] \emph{Degenerate-maps.} An elliptical conic is said to be degenerate when the intersection between the cone and the plane is a line, namely the plane is tangent to the cone. In these case the map $\tA$ sends $\Extr(\Stset)$ into an extremal ray of $\Stset_+$.
   \item[{\bf c.}] \emph{Strictly elliptical-maps.} In these case $\Extr(\Stset)$ is mapped into a true ellipse.  
\end{enumerate}
First notice that it is $\set{Aut}(\Stset)=\grp O(2)$, namely the local automorphisms of the model
are the rotation $\tR_\phi$ around the cone axis plus the reflections $\tS_\phi$ through the axis at
$\phi$. The elliptical-maps correspond to the transformations
$\tR_\phi\circ\tA^\gamma\circ\transp{\tR_\theta}$,
$\tS_\phi\circ\tA^\gamma\circ\transp{\tS_\theta}$, $\tR_\phi\circ\tA^\gamma\circ\transp{\tS_\theta}$
and $\tS_\phi\circ\tA^\gamma\circ\transp{\tR_\theta}$ where $\tA^\gamma$ is the transformation
having the following Bloch representative
\begin{equation}\label{e:ell_gen}
\mat A^\gamma=\footnotesize\begin{bmatrix}
\gamma&0&(1-\gamma)\\0&\sqrt{2\gamma -1}&0\\
(1-\gamma)&0&\gamma\end{bmatrix}\normalsize,\qquad
\gamma\in\left[\tfrac{1}{2},1\right].
\end{equation}
For example the maps corresponding to $\tS_\phi\circ\tA^\gamma\circ\transp{\tS_\theta}$ are 
\begin{equation}\label{e:ell_Tr}
\begin{split}
&\qquad\qquad\qquad\qquad\qquad\qquad\quad\mat S_\phi\mat A^\gamma\transp{\mat S_\theta}=\\
&\scriptsize\begin{bmatrix}\gamma\cos 2\theta\cos 2\phi+\sqrt{2\gamma-1}\sin 2\theta\sin 2\phi&
\gamma\cos 2\theta\sin 2\phi-\sqrt{2\gamma-1}\sin 2\theta\cos 2\phi&(1-\gamma)\cos 2\theta\\
\gamma\sin 2\theta\cos 2\phi-\sqrt{2\gamma-1}\cos 2\theta\sin 2\phi&
\gamma\sin 2\theta\sin 2\phi+\sqrt{2\gamma-1}\cos 2\theta\cos 2\phi&(1-\gamma)\sin 2\theta\\
(1-\gamma)\cos 2\phi&(1-\gamma)\sin 2\phi&\gamma\end{bmatrix}\normalsize\\
&\qquad\qquad\qquad\qquad\qquad \phi,\theta\in(0,\pi],\;\gamma\in\left[\tfrac{1}{2},1\right],
\end{split}
\end{equation}
\normalsize while the other three combinations, $\tR_\phi\circ\tA^\gamma\circ\transp{\tR_\theta}$,
$\tR_\phi\circ\tA^\gamma\circ\transp{\tS_\theta}$ and
$\tS_\phi\circ\tA^\gamma\circ\transp{\tR_\theta}$, are exactly the same a part from signs. Clearly,
$\Erays(\Trnset_+)$ is made of all the maps proportional to the above ones.  According to the value
of the parameter $\gamma$ it is possible to identify the following three different kinds of maps.
\par {\bf a.}  For $\gamma=1$ we achieve the circular-maps. It is easy to check that these maps are
exactly the rotations and the reflections, namely the local automorphisms of the model
$\set{Aut}(\Stset)$. Accordingly, the last row of their Bloch representatives is the deterministic
effect $\mat\lambda(e)=[0,0,1]$.
\par{\bf b.}  For $\gamma=1/2$ we achieve the degenerate-maps. Denoting by $a_\phi$, with
$\phi\in(0,2\pi]$, the extremal effects lying on the circle at $z=1/2$ in the left Fig.
\ref{fig:TC_cones_sd}, these maps are exactly $\Extr(a_\phi)$ for $\phi\in(0,2\pi]$. Consider for
example the effect $a_0$ having representative $\mat\lambda(a_0)=(1/2,0,1/2)$. According to Bloch
representation, the extremal map in Eq. (\ref{e:ell_gen}) (for $\lambda=1/2$) has effect $a_0$. All
the extremal maps having this effect, namely $\Extr(a_0)$, are achieved from the previous one by
applying $\set{Aut}(\Stset)$ on the left. From Proposition \ref{prop:Bloch_repr3} we know that
$\{\Extr(a_\Phi),\,\phi\in(0,2\pi]\}=\Extr(a_0)\circ\set{Aut}(\Stset)$, namely all maps are obtained
by applying automorphisms on the right of the maps in $\Extr(a_0)$.
\par{\bf c.} For $\gamma\in(1/2,1)$ we get the strictly elliptical-maps. These transformations
belong to the non extremal effects (equivalence classes) whose Bloch representatives are the vectors
$[(1-\gamma)\cos\phi,(1-\gamma)\sin\phi,\gamma]$, for $\gamma\in(1/2,1)$ and $\phi\in(0,2\pi]$.
According to Observation \ref{obs:1} in this model, as in Quantum Mechanics, there exist extremal
transformations having non extremal effects.

\subsection{The bipartite cone of states}\label{ss:TC_bip_states} We know that the isomorphism $\Trnset_+(\rA)\simeq\Stset_+(\rA\rA)$ induced by the chosen faithful state leads to the relation in Eq. (\ref{e:expl_isom_TC}) between bipartite states and physical transformations. Then the same matrixes representing the extremal maps $\tR_\phi\circ\tA^\gamma\circ\transp{\tR_\theta}$,
$\tS_\phi\circ\tA^\gamma\circ\transp{\tS_\theta}$, $\tR_\phi\circ\tA^\gamma\circ\transp{\tS_\theta}$ and
$\tS_\phi\circ\tA^\gamma\circ\transp{\tR_\theta}$  
represent all the pure bipartite states too (apart from normalization).
For completeness we report explicitly the matrices representing the normalized states associated to the transformations $\tS_\phi\circ\tA\circ\transp{\tS_\theta}$
\begin{equation}\label{e:TC_pure_bip_states}
\begin{split}
&\,\Psi=\frac{(\tS_\theta\otimes\tS_\phi\circ\tA^\gamma)\Phi}{\Phi(e,a\circ\tS_\theta)}
\Rightarrow\mat\Psi=\frac{\mat\Phi\transp{(\mat S_\phi\mat A^\gamma\transp{\mat S_\theta})}}{\transp{\mat\lambda(e)}\mat\Phi\mat\lambda(a\circ\tS_\theta)}=
\frac{\transp{(\mat S_\phi\mat A^\gamma\transp{\mat S_\theta})}}{\gamma}=\\
&\scriptsize\begin{bmatrix}
\cos 2\theta\cos 2\phi+\frac{\sqrt{2\lambda-1}}{\lambda}\sin 2\theta\sin 2\phi&
\sin 2\theta\cos 2\phi-\frac{\sqrt{2\lambda-1}}{\lambda}\cos 2\theta\sin 2\phi&
\frac{(1-\lambda)}{\lambda}\cos 2\phi\\
\cos 2\theta\sin 2\phi-\frac{\sqrt{2\lambda-1}}{\lambda}\sin 2\theta\cos 2\phi&
\sin 2\theta\sin 2\phi+\frac{\sqrt{2\lambda-1}}{\lambda}\cos 2\theta\cos 2\phi&
\frac{(1-\lambda)}{\lambda}\sin 2\phi\\
\frac{(1-\lambda)}{\lambda}\cos 2\theta&\frac{(1-\lambda)}{\lambda}\sin 2\theta&1
\end{bmatrix}\normalsize\\
&\qquad\qquad\qquad\qquad\phi,\theta\in(0,\pi],\;\gamma\in\left[\tfrac{1}{2},1\right].
\end{split}
\end{equation}
Notice that the states associated to the degenerate-maps, are the \emph{factorized bipartite pure
  states} given by $\mat l(\omega_\theta)\otimes\mat l(\omega_\phi)$, $\forall
\omega_\theta,\omega_\phi\in\Extr(\Stset)$. The states corresponding to the circular and strictly
elliptical-maps are the \emph{non-local bipartite pure states} of the model. In particular, as will
be investigated in a forthcoming publication, the states associated to local automorphisms
$\set{Aut}(\Stset)$ achieve the Cirel'son bound (see Ref. \cite{Cirel}) of the model.\footnote{The
  Cirel'son bound of the two-clock model is the same of the Quantum Mechanics one, namely
  $2\sqrt{2}$.}

\subsection{Purifiability at the single system level}\label{ss:TC_purifiability}
Differently from the Popescu-Rohrlich probabilistic model the two-clock world satisfies Postulate
PURIFY at the single system level as stated in the following Proposition.
\begin{proposition}\label{prop:TC_purification}
In the two-clock model with the cones $\Trnset_+$ and $\Stset_+(\rA\rA)$ respectively introduced
in Subsecs. \ref{ss:TC_transf} and \ref{ss:TC_bip_states}, any mixed local state has purification
unique up to local automorphisms on the purifying system. 
\end{proposition}
\Proof Notice that in Bloch representation the marginalization on the purifying system of a
bipartite state is simply the last column of its representative matrix. Consider then the pure
bipartite states in Eq. (\ref{e:TC_pure_bip_states}). By taking the marginal over the purifying
system (the second one) we get the set of local states
\begin{equation}
\left\{\;\Psi(\cdot,e)\equiv\mat\Psi\mat\lambda(e)=\begin{bmatrix}\frac{(1-\gamma)}{\gamma}\cos\phi\\
\frac{(1-\gamma)}{\gamma}\sin\phi\\1\end{bmatrix},\;
\phi\in(0,2\pi],\;\gamma\in\left[\tfrac{1}{2},1\right]\;\right\}=\Stset,
\end{equation}
which coincide with the whole set of states $\Stset$, proving purifiabilility of the model. The
uniqueness up to local automorphisms is easily verified. In fact, first notice that if $\Psi$ is a
purification of $\omega$, \ie $\Psi(\cdot,e)=\omega$, then also the states $(\tI\otimes\tR_\phi)\Psi$ and
$(\tI\otimes\tS_\phi)\Psi$ are purifications of $\omega$, because the last column of their representative
matrixes is the same of the $\mat\Psi$'s one. Then, suppose that there exists another purification
of $\omega$---say $\Psi^\prime$--- which is not connected to $\Psi$ by a local automorphism acting
on the second system. But, according to the pure bipartite states introduced in Subsec.
\ref{ss:TC_bip_states}, there exist $\tD_1,\tD_2\in\set{Aut}(\Stset)$ such that
$\Psi^\prime=(\tD_1\otimes\tD_2)\Psi$ and then $\Psi^\prime=(\tI\otimes\tD_2\transp{\tD_1})\Psi$ which
contradicts the hypothesis.\qed

\subsection{Exploring teleportation and purifiability}
The probabilistic model introduced in this section does not allow teleportation, because the inverse
of the preparationally faithfull state is not a true bipartite effect. In fact considering the state
$\Psi=(\tI\otimes\tR_{\pi})\in\Stset_+(\rA\rA)$ we get $\Phi^{-1}(\Psi)=-1$, which is negative. More
precisely we get
\begin{equation}\label{e:TC_tel}
\Phi^{-1}(\Psi)\leq 0\qquad\forall\Psi=(\tI\otimes\tR_{\phi})\Phi\;\text{ with }\;\phi\in[5\pi/6,7\pi/6].
\end{equation}
Thus Postulate FAITHE does not hold in this model and according to Corollary \ref{cor:FAITH_tel_cor}
teleportation is not achievable.  A good question is how the set $\mathfrak T_+$, and then
$\Stset_+(\rA\rA)$, has to be restricted in order to achieve a theory which allows teleportation
preserving the purifiability of the theory. Indeed, reducing the set of physical transformations we
also reduce $\Stset_+(\rA\rA)$ and, by duality, the set of bipartite effects $\Cntset_+(\rA\rA)$
grows.
\begin{observation}One may try to get a theory with teleportation excluding some automorphisms from
  $\Trnset$. Indeed excluding rotations in $\grp O(2)$, the states $\Psi$ in Eq. (\ref{e:TC_tel})
  are no longer states of the theory and $\Phi^{-1}$ would be a true effect. On the other hand we
  cannot take reflections as the only physical automorphisms because $\Trnset$ is closed under
  combination and all rotations are achievable by composing two reflections. We could eventually
  reduce the set of physical automorphisms to $\grp {{SO}}(2)$ but obviously teleportation would be
  still impossible.
\end{observation}
\begin{observation}As in the two-box world Proposition \ref{prop:super_faith} ensures that also the
  two-clock wold does not admit a \emph{super-faithful} state.
\end{observation}  
In the following we will us the abbreviation {\em purifiability of states}, to express the existence
of purification of states, also uniquely up to reversible channels on the purifying system.

From the impossibility of achieving teleportation in the present model follows an interesting
property of the probabilistic theories in general.
\begin{proposition} In a probabilistic theory, purifiability of single-system states does not imply
  purifiability at higher multipartite levels of the theory.
\end{proposition}
\Proof The proof of this statement is simply the counterexample given by the two-clock model
constructed in this Section. In fact from Proposition \ref{prop:TC_purification} we know that the
model allows a purification for every mixed local state, unique up to reversible channel on the
purifying system. This means that uniqueness of purification holds at the single-system level. On
the other hand, according to Proposition \ref{prop:entsw}, the same property at all the multipartite
levels of the theory should imply the possibility of achieving probabilistic teleportation, which has
been already excluded.\qed

\subsection{A global feature from the local system structure}
Here we observe a global feature of the two-clock probabilistic theories arising from the shape of
the local cones.
\begin{proposition}It is impossible to construct a probabilistic theory having a disk as local set of states and a self-dual bipartite system at the same time.
\end{proposition}
\Proof The model constructed in this Section is self-dual at the single system level as
geometrically represented in the left Fig. \ref{fig:TC_cones_sd}. From the local self-duality it
follows that the bipartite system is self-dual in correspondence of its ``local component'', namely
the factorized bipartite states $\omega_1\otimes\omega_2$, with $\omega_1,\omega_2\in\Stset$, are
proportional to bipartite effects ($a_1\otimes a_2$ with $\omega_1=\Phi(a_1,\cdot)$ and
$\omega_2=\Phi(a_2,\cdot)$). On the other hand the bipartite system is not self-dual because of its
``non-local component''. Indeed not all the bipartite states associated, by the faithful state
$\Phi=\sum_{i=1}^{3}\lambda_i\otimes\lambda_i$, to the local automorphisms $\set{Aut}(\Stset)$ are
proportional to bipartite effects. Regarding the states $\Psi_\phi=(\tI\otimes\tR_\phi)\Phi$ as bipartite
functionals over $\Stset(\rA\rA)$ we get, for example, $\Psi_{2\pi}(\Psi_\pi)<0$, namely
$\Psi_{2\pi}$ is not proportional to a bipartite effect.
\par The only way to make the bipartite states associated by the faithful state to the local automorphisms $\set{Aut}(\Stset)$ proportional to bipartite effects is to modify the faithful state of the theory. To achieve this goal the faithful state must be $\Phi^\prime=\frac{1}{\sqrt{2}}(\lambda_1\otimes\lambda_1+\lambda_2\otimes\lambda_2)+
\lambda_3\otimes\lambda_3$. We know that the faithful state induces also the isomorphism $\Phi^\prime(a,\cdot)=\omega_a$  between the local cones of effects and states. Differently from the old faithful state $\Phi$, the new one squeezes the local cone of states with respect to the cone of effects, as showed in the right Fig.
\ref{fig:TC_cones_sd}, destroying the local self-duality of the model. Naturally a model without local self-duality cannot be seldual at the bipartite system level because of its factorized component.\qed 

\section{An hidden quantum model for the two-clock world: the rebit}\label{s:hiddqm}
In the class of probabilistic theories having a disk as local convex set of states a special case is
that of the \emph{equatorial qubit}. In fact, the convex set of \emph{qubit} states is the
$3$-dimensional ball known as \emph{Bloch sphere}, and the clock corresponds to the qubit in the
equatorial plane. This model is also called {\bf rebit}, where ``re'' stays for real, and
corresponds to Quantum Mechanics on a two-dimensional real Hilbert space.  The peculiarity of the
rebit model is that it violates \emph{local observability}.

\subsection{Local states and effects}
Consider as usual the canonical basis $\mat l=\{l_i\}$ and $\mat \lambda=\{\lambda_i\}$ with $i=1,2,3$ for
$\Cntset_\Reals$ and $\Stset_\Reals$ embedded into $\Reals^3$ as Euclidean spaces. Inspired by the well known qubit model, upon defining the operator vector $\sigma=[\sigma_z,\sigma_x,I]$, and introducing the canonical orthonormal basis $\{u_j\}$ for $\Reals^3$, we define the following bijective map
\begin{equation}\label{e:qntmap}
\qnt: r\in\Reals^3\leftrightarrow\qnt( r)\in\Her(\Reals^2),\quad
\begin{cases}\qnt(r)=\tfrac{1}{\sqrt2} r\cdot\sigma,&\\
\qnt^{-1}(A)=\tfrac{1}{\sqrt2}\Tr[A\sigma]\cdot u,&
\end{cases}
\end{equation}
where $u$ is the vector having the $\Reals^3$ basis vectors $\{u_i\}$ as components.
We get the pairing relation\footnote{One has:
$\qnt^{-1}(A)\cdot\qnt^{-1}(B)=\tfrac{1}{2}\Tr[A\sigma]\cdot\Tr[B\sigma]=
\Tr[(A\otimes B)\tfrac{1}{2}\sum_i\sigma_i\otimes\sigma_i]=\Tr[A\transp{B}]-\Tr[(A\otimes
B)\sigma_y\otimes\sigma_y]=\Tr[AB]$, where we have subtracted the component concerning $\sigma_y$.}
\begin{equation}\label{e:bellet_prod}
\qnt(r)\bullet\qnt(s):=\Tr[\qnt(r)\qnt(s)], \quad\Tr[AB]=\qnt^{-1}(A)\cdot\qnt^{-1}(B).
\end{equation}
The symbol $\bullet$ denotes a ``scalar product" between elements in $\Her(\Reals^2)$ as defined in the last equation, and it is easy to verify that
$\qnt(r)\bullet\qnt(s)=r\cdot s$, $\forall r,s\in\Reals^3$. In terms of the canonical basis one has
\begin{equation}
\qnt(u_i)=\qnt(l_i)=\qnt(\lambda_i)=\tfrac{1}{\sqrt2}\sigma_i,\quad (l_j,\lambda_i)=
\tfrac{1}{2}\qnt^{-1}(\sigma_i)\cdot\qnt^{-1}(\sigma_j).
\end{equation}

Specializing the map to states and effects of a clock we have the states and effects of the rebit (the hidden quantum model)
\begin{equation}
\omega\in\Stset,\;\rho=\tfrac{1}{\sqrt2}\qnt(\mat l(\omega))\in\set{St}(\Reals^2),\quad
a\in\Cntset,\;A=\sqrt2\qnt(\mat\lambda(a))\in\Bndp{\Reals^2},
\end{equation}
with Born rule
\begin{equation}
\Tr[A\rho]=\qnt(\mat l(\omega))\bullet\qnt(\mat\lambda(a))\equiv(a,\omega),
\end{equation}
$\set{St}(\Reals^2)$ denoting the set of symmetric real matrices with unit trace. Notice that it is $\rho=\tfrac{1}{\sqrt{2}}\big(I+\mat{\hat l}(\omega)\cdot\sigma\big)$,
where $\mat{\hat l}(\omega)$ is the \emph{Bloch vector} representing the point in the disk of states $\Stset$.
  The extension of
the map to tensor product is given by the ``commutation rule''
$\qnt\otimes=\otimes\qnt$, namely
\begin{equation}\label{e:comm}
\qnt(r\otimes s):=\qnt(r)\otimes\qnt(s),\quad
\qnt^{-1}(A\otimes B)=\qnt^{-1}(A)\otimes\qnt^{-1}(B).
\end{equation}
In the following we will use the abbreviate notation $\qnt(\omega):=\qnt(\mat l(\omega))$ for states and
$\qnt(a):=\qnt(\mat\lambda(a))$ for effects.

\subsection{The bipartite system: states and transformations}
The faithful state is the bipartite functional $\Phi$ such that
$\Phi(l_i,l_j)=\delta_{ij}$, whence the corresponding operator is given by
\begin{equation}
\tfrac{1}{2}\qnt(\Phi)=\tfrac{1}{2}\sum_{i=1}^3\qnt(\lambda_i)\otimes\qnt(\mat\lambda_i)=
\tfrac{1}{4}(I\otimes I+\sigma_x\otimes\sigma_x+\sigma_z\otimes\sigma_z),
\end{equation}
which is an Hermitian (non positive) operator with unit trace. Notice that such operator differs
from the quantum maximally entangled state
\begin{equation}\label{e:max_ent_state}
\tfrac{1}{2}|I\kk\bb
I|=\tfrac{1}{4}(I\otimes I+\sigma_x\otimes\sigma_x+\sigma_y\otimes\sigma_y+\sigma_z\otimes\sigma_z), 
\end{equation}
by the term $\frac{1}{4}\sigma_y\otimes\sigma_y\not\in\Bnd{\Reals^2}\otimes\Bnd{\Reals^2}$.
The term $\sigma_y\otimes\sigma_y\in\Bnd{\Reals^4}$ corresponds to the null linear form $\Xi$ over
$\Reals^3\otimes\Reals^3$ given by
\begin{equation}
\Xi(R)=\Tr[\sigma_y\otimes\sigma_y\qnt(R)]=0,\qquad\forall R\in\Reals^3\otimes\Reals^3.
\end{equation}
Notice that the transposition acts as the identity map over $\qnt(\Reals^3)$, since transposition
leaves $\sigma_x$, $\sigma_z$ and $I$ invariant, whence $\qnt^{-1}[\transp{\qnt(a)}]=\tI(a)$. Using this identity one can also see that the maximally entangled state is another equivalent
representation of the faithful state $\Phi$, since $\forall r,s\in\Reals^3$ one has
\begin{equation}
\tfrac{1}{2}\bb I|\qnt(\mat\lambda(a))\otimes\qnt(\mat\lambda(b))|I\kk=
\tfrac{1}{2}\Tr[\qnt(\mat\lambda(a))\transp{\qnt(\mat\lambda(b))}]=
\mat\lambda(a)\cdot\mat\lambda(b)=\Phi(a,b),
\end{equation}
(transposition works as the identity over $\sigma_x$, $\sigma_z$ and $I$).
\par Let's now represent maps in the hidden quantum model. A generic bipartite state is represented as
\begin{equation}
\Psi=\sum_{ij}\Psi_{ij}\lambda_i\otimes\lambda_j,\quad \Psi_{ij}=\Psi(l_i,l_j),
\end{equation}
and the local action of the transformation $\tA$ is given by
\begin{equation}
\begin{split}
&(\tA\otimes\tI)\Psi(l_i,l_j)=\Psi(l_i\circ\tA,l_j)=\sum_kA_{ik}\Psi(l_k,l_j)=\\
\sum_{nklm}&A_{nk}\Psi_{lm}\lambda_n(l_i)l_k(\lambda_l)\lambda_m(l_j)
=\tfrac{1}{2}\Tr[\qnt(\tA)\star\qnt(\Psi)\sigma_i\otimes\sigma_j],
\end{split}
\end{equation}
where
\begin{equation}
\qnt(\tA):=\tfrac{1}{2}\sum_{nk}A_{nk}\sigma_n\otimes\sigma_k,\qquad
\qnt(\Psi):=\tfrac{1}{2}\sum_{lm}\Psi_{lm}\sigma_l\otimes\sigma_m,
\end{equation}
and
\begin{equation}
A\star B=\Tr_2[(A\otimes I)(I\otimes B)].
\end{equation}
The algebra of transformations allows a representation as operator algebra over $\Her(\Reals^2)$ and denoting by $\qA$ $(\tilde\qA)$ and $\qI$ the operators corresponding respectively to $\tA$ $(\tA^\prime)$ and $\tI$ one has
\begin{equation}
(\tA\otimes\tI)\Psi(l_i,l_j)=\tfrac{1}{2}\Tr[(\qA\otimes\qI)\qnt(\Psi)\sigma_i\otimes\sigma_j]
=\tfrac{1}{2}\Tr[\qnt(\Psi)\tilde\qA(\sigma_i)\otimes\sigma_j],
\end{equation}
whence
\begin{equation}
\qnt[(\tA\otimes\tI)\Psi]=\qnt(\tA)\star\qnt(\Psi)=(\qA\otimes\qI)\qnt(\Psi)=
\qnt(\Psi)(\tilde\qA\otimes\qI).
\end{equation}
Now we have to choose the physical transformations of the model. In the previous two clocks models $\Stset$ was a $2$-dimensional convex set. Then was  $\dim(\Stset_\Reals)=3$ and $\dim(\Trnset_\Reals)=9$. The set $\Trnset^{q}_\Reals$ for the qubit model is the linear $\Span$ of the quantum operations $\sigma_i\cdot\sigma_j$, for $i,j=1,2,3,4$ and then
\begin{equation}\label{e:span_qbit}
\Trnset_{\Reals}^{q}=\Span\left\{\tA_{11},\tA_{22},\tA_{33},\tA_{44},
\tA_{12},\tA_{13},\tA_{23},
0\tA_{14},\tA_{24},\tA_{34}\right\}
\end{equation}
where
\begin{equation}\label{e:quant_oper}
\tA_{ij}\omega:=\qnt^{-1}[\sigma_i\qnt(\omega)\sigma_j],\qquad
\forall\omega\in\Stset.
\end{equation}
Notice that $\dim(\Trnset^{q}_\Reals)=10$\footnote{The qubit model is based on the $2$-dimensional Hilbert space $\sH$ and $\dim(\Stset)=4$ where $\Stset=\sS(\sH)$ is the states space. According to the Choi-Jamiolkowski isomorphism, in the $16$-dimensional linear space $\Bnd{\sH\otimes\sH}$ we take only the operator corresponding to completely positive maps and we get $\dim(\Trnset_\Reals^q)=10$ (the only Hermitian matrices are allowed).}.
Here we are considering the equatorial qubit (rebit)
and the space $\Trnset_\Reals\equiv\Lin{\Cntset_\Reals}=\Lin{\Reals^3}$ of linear maps over $\Reals^3$ can be obtained from the one in Eq. (\ref{e:span_qbit}) as follows\footnote{The symbols $\Re$ and $\Im$ stay respectively for Real and Imaginary part.}
\begin{equation}\label{e:trnsetclock}
\Trnset_\Reals^r=\Span\left\{\tA_{11},\tA_{22},\tA_{33},\tA_{44},\Re\tA_{12},\Re\tA_{13},\Re\tA_{23},
\Im\tA_{14},\Im\tA_{24},\Im\tA_{34}\right\}
\end{equation}
with $\tA_{ij}$ as in Eq. (\ref{e:quant_oper}).
\par We know that the automorphisms of the convex set of states $\Stset$ are given by the rotations
$\tR_\phi,\;\phi\in[0,2\pi)$ along with the reflections $\tS_\phi,\;\phi\in[0,\pi)$ through the axis
at $\phi$. Therefore is $\set{Aut}(\Stset)=\grp O(2)$. Taking the physical maps as in Eq.
(\ref{e:trnsetclock}) we get all rotations and reflections of the disk of states. In fact the
quantum operations achieve the automorphisms of the qubit system namely the rotations in
$\grp{{SO}}(3)$. On the other hand the rotations of a sphere include not only the rotations of its
equatorial disk but also its reflections.

\subsection{Ghosts}
As already mentioned the set of transformations in the hidden quantum model should have dimension 10 from the qubit quantum operations. On the other hand not all the matrices representing the 10 independent quantum operations are linearly independent when applied to the rebit. In fact the completely positive maps
\begin{equation}\label{e:two_tr}
\sigma_x\cdot\sigma_x+\sigma_z\cdot\sigma_z-I\cdot I,\qquad\quad\sigma_y\cdot\sigma_y,
\end{equation}
are not distinguishable by their local action over a rebit. As can be easily verified, the matrixes
representing the quantum operations in Eq. (\ref{e:two_tr}), which are locally distinguishable on a
qubit, become the same when we take the ``Quantum Mechanics of real matrixes''. Clearly, by
identification of locally indistinguishable transformations (namely taking the space of
transformations having dimension 9), the local observability principle is satisfied. This is not the
case if the space of transformations is in dimension 10. In fact in that case there exists two
transformations indistinguishable by local tests but discriminable by bipartite measurements.
  
\subsection{Bipartite effects and teleportation}
In Eq. (\ref{e:bellet_prod}) we have defined the product
$\qnt(r)\bullet\qnt(s):=\Tr[\qnt(r)\qnt(s)]$, $\forall r,s\in\Reals^3$, from which the local states
effects pairing relation $\qnt(\omega)\bullet\qnt(a)\equiv(a,\omega)$. We can coherently extend the
product $\bullet$ as follows
\begin{equation}\label{sc_pr_gen}
\qnt(R)\bullet\qnt(S)=\tfrac{1}{2}\sum_{i,j}\Tr[\qnt(R)\star\qnt(S)\sigma_i\otimes\sigma_j]
\qquad\forall R,S\in\Reals^3\otimes\Reals^3,
\end{equation}
to represent the pairing relation between bipartite states and effects as
\begin{equation}\label{e:pairbip}
\qnt(E)\bullet\qnt(\Psi)=(E,\Psi).
\end{equation}
\begin{proposition}The rebit model does not allow probabilistic teleportation, nor a superfaithful
  state.
\end{proposition}
\Proof Let's first take the generalized effect corresponding to the
inverse matrix of $\Phi$, \ie which would achieve teleportation, and let see if it is a true effect.
The matrix multiplication between two (considering $\Phi^{-1}$ as a map) must be as follows
\begin{equation}
\delta_{ij}=\tfrac{1}{2}\Tr[\qnt(\Phi^{-1})\star\qnt(\Phi)\sigma_i\otimes\sigma_j],
\end{equation}
and taking $F^\Phi=\alpha\Phi^{-1}$ we get
\begin{equation}
(F^\Phi,\Phi)=\alpha(\Phi^{-1},\Phi)=\alpha\qnt(\Phi^{-1})\bullet\qnt(\Phi)=3\alpha,
\end{equation}
whence $\alpha=1/3$, and one would have probability of successful teleportation
\begin{equation}
F^\Phi\omega\Phi(e)=(F^\Phi,e)(\omega,\Phi)=\alpha\Tr[(\qnt(\Phi^{-1})\otimes \tfrac{1}{\sqrt2}I)(\qnt(\omega)\otimes\qnt(\Phi))]=\tfrac{1}{3}.
\end{equation}
On the other hand $F^\Phi$ is not a true effect. Consider the state $\Psi$ given by
\begin{equation}
\Psi=(\tA_{44}\otimes\tI)\Phi=
\qnt^{-1}[(\sigma_4\otimes\qI)\qnt(\Phi)(\sigma_4\otimes\qI)]
\end{equation}
where $\sigma_4\cdot\sigma_4$ is a completely positive map and then a physical transformation. Explicitly is 
\begin{equation}
\begin{split}
\tfrac{1}{2}\qnt(\Psi)=&\tfrac{1}{2}\qnt(\tA_{44})\star\qnt(\Phi)=
\tfrac{1}{2}(\sigma_4\otimes\qI)\qnt(\Phi)(\sigma_4\otimes\qI)\\
=&\tfrac{1}{4}(I\otimes I -\sigma_x\otimes\sigma_x-\sigma_z\otimes\sigma_z).
\end{split}
\end{equation}
Considering $(F^\Phi,\Psi)$,
\begin{equation}
(F^\Phi,\Psi)=\qnt(F^\Phi)\bullet\qnt(\Psi)=
\tfrac{1}{3}\qnt(\Phi^{-1})\bullet\qnt(\Psi)=-1,
\end{equation}
we find a negative value meaning that $F^{\Phi}$ is not a bipartite effect. Postulate FAITHE is not satisfied and according to Corollary \ref{cor:FAITH_tel_cor} teleportation is not achievable.  Moreover, from Proposition \ref{prop:super_faith}, the rebit probabilistic theory does not allow a super-faithful state.\qed
\subsection{Purifiability}
It is well known that the Quantum Meechanics of real matrices satisfies Postulate PURIFY. For each local state $\qnt(\omega)=\rho_\omega=(I+\mat{\hat l}(\omega)\cdot\sigma)/\sqrt{2}$ of the rebit system,
we find a pure bipartite state $|\,{\rho_\omega}^{1/2}\kk$ which purifies it.

The bipartite state $\qnt(\Psi)$ corresponding to $|\,{\rho_\omega}^{1/2}\kk$ is given by the
relation
\begin{equation}
\tfrac{1}{2}\qnt(\Psi)=|\,{\rho_\omega}^{1/2}\kk\bb{\rho_\omega}^{1/2}|
\quad\text{with}\quad 
\Tr_2\left[|\,{\rho_\omega}^{1/2}\kk\bb{\rho_\omega}^{1/2}|\right]=\rho_\omega.
\end{equation}
All the purifications of a state are connected by local automorphisms on the purifying system, that is 
$\B{e}_2(\tI\otimes\tD)\K{\Psi}=\K{\omega}_1$ $\forall\tD\in\set{Aut}(\Stset)$, or in quantum notation,
\begin{equation}
\Tr_2\left[(\qI\otimes\qD)|\,{\rho_\omega}^{1/2}\kk\bb{\rho_\omega}^{1/2}
|(\qI\otimes\tilde\qD)\right]=\rho_\omega.
\end{equation}
In the last equation we have used the relation $\qD\tilde\qD=\qI$.

We have already shown that FAITHE is not satisfied. Therefore, from Proposition \ref{prop:entsw}, the uniqueness of purification, up to reversible channels, at all the multipartite levels, is not satisfied.

\section{Toy-theory 3: the two-Spin-factor world}
The convex set of states of the clock is the disk $\Stset=\mathbb B^2$ while for the qubit one has
$\Stset=\mathbb B^3$. Therefore it seems interesting to investigate probabilistic theories with
$\Stset=\mathbb B^n$. The local system of these theories is denoted {\bf
  $\mathbf{(n)}$spin-factor}. Naturally, as noticed for the clocks world, many probabilistic
theories may have the same $(n)$spin-factor as local system. 
\subsection{The self-dual $\mathbf{(n)}$spin-factor, its states and effects}
Consider the self-dual $(n)$spin-factor and denote as usual by $\mat l=\{l_i\}$ and
$\mat\lambda=\{\lambda_j\}$, with $i,j=1,\ldots,n+1$, the canonical basis for $\Stset_\Reals$ and
$\Cntset_\Reals$. The cones of states and effects coincide, whence
\begin{equation}
\Stset_+=\left\{\mat l(\omega)\;|
\;x_1^2+\ldots+x_n^2\leq x_{n+1}^2\right\},\quad\Cntset_+=\left\{\mat\lambda(a)\;|
\;x_1^2+\ldots+x_n^2\leq x_{n+1}^2\right\}.
\end{equation}
Naturally the set of states is the section of the cone at $x_{n+1}=1$, while its truncation, from
the order relation $0\leq a\leq e$, gives the set of effects
\begin{equation}\label{e:sf_states}
\begin{split}
\Stset&=\left\{\mat l(\omega)\;|
\;x_1^2+\ldots+x_n^2\leq 1\right\},\\
\Cntset&=\left\{\mat\lambda(a)\;|\;
\;x_1^2+\ldots+x_n^2\leq \min\left(x_{n+1}^2,(1-x_{n+1})^2\right),\;x_{n+1}\in[0,1]\right\}.
\end{split}
\end{equation}

\subsection{Wath is special about the $\mathbf{(3)}$spin-factor?}
As for the clocks---the $(2)$spin-factors---the probabilistic theory is defined only at the
single-system level. Therefore we need to extend the theory at the bipartite level. We do this by
assuming a faithful state that is the $(n+1)$-dimensional generalization of the one given in Eq.
(\ref{e:TC_faith_state}), namely the bipartite functional
\begin{equation}\label{e:sf_faith_state}
\Phi=\sum_{i=1}^{n+1}\lambda_i\otimes\lambda_i.
\end{equation}
Such state, being represented by the identical matrix $\mat\Phi=\mat I_{n+1}$, realizes the
cone-isomorphism $\Stset_+\simeq\Cntset_+$ via the map $\omega_a:=\Phi(a,\cdot)=a$. In our
probabilistic framework, from the isomorphism $\Stset_+(\rA\rA)\simeq\mathfrak T_+$ given by
\begin{equation}
\Psi=(\tI\otimes\tA)\Phi\Rightarrow\mat\Psi=\transp{\mat A},
\end{equation}
the cone of bipartite states $\Stset_+(\rA\rA)$ can be generated from the set $\mathfrak T_+$ of
two-positive maps (the physical transformations of our model), while the bipartite set of effects
$\Cntset_+(\rA\rA)$ follows by duality from $\Stset_+(\rA\rA)$.
\par The analysis of the spin-factors probabilistic world is extremely technical and in this section
we only give an interesting result. First notice that for an $(n)$spin-factor is
$\set{Aut}(\Stset)=\grp O(n)$. Therefore the following proposition holds
\begin{proposition}\label{prop:O(n)} Consider a probabilistic theory having an $(n)$spin-factor as
  local system with $\set{Aut}(\Stset)\in\Trnset$. Then, for each $n$, Postulate FAITHE is not
  satisfied.
\end{proposition}
\Proof If the whole set $\set{Aut}(\Stset)=\grp O(n)$ is physical it is always possible to find a
bipartite state $\Psi\in\Stset(\rA\rA)$ such that $F(\Psi)<0$, where $F=\alpha\Phi^{-1}$ is the
bipartite functional inverting the faithful state $\Phi$ (from Eq. (\ref{e:sf_faith_state}) is
$\Phi^{-1}=\sum_{i=1}^{n+1}l_i\otimes l_i$ and $\mat\Phi^{-1}=\mat I_{n+1}$). In fact, consider the
automorphism $\tD\in\grp O(n)$ reversing the direction of every vector. The $(n+1)\times(n+1)$
matrix $\mat D$ representing $\tD$ in our basis is the diagonal matrix with $D_{ii}=-1$ for
$i=1,\ldots,n$ and $D_{n+1,n+1}=1$. Therefore the state
$\Psi=(\tI\otimes\tD)\Phi=-\sum_{i=1}^{n}(\lambda_i\otimes\lambda_i) +\lambda_{n+1}\otimes\lambda_{n+1}$
achieves
\begin{equation}
F(\Psi)=-\sum_{i,j=1}^{n}l_i(\lambda_j)\otimes l_i(\lambda_j)+l_{n+1}(\lambda_{n+1})\otimes l_{n+1}(\lambda_{n+1})<0\quad\forall n\geq 2.
\end{equation}    
In general the automorphism $\tD$ is a combination of reflections and rotations and it is not the
only combination achieving a state $\Psi$ with $F(\Psi)<0$.\qed In regard of the closure of
$\Trnset$ under composition of transformations, it is possible to reduce the set of physical
automorphisms from $\grp O(n)$ to its subgroup $\grp {{SO}}(n)$, that is the component connected to
the identical transformation. On the other hand the following proposition holds:
\begin{proposition} Consider a probabilistic theory having as local system an $(n)$spin-factor with
  physical automorphisms given by the group $\grp {{SO}}(n)$. For each $n\neq 3$ FAITHE is still
  violated.
\end{proposition}   
\Proof
It is easy to see that
\begin{equation}
\forall n\neq 3\qquad\exists\tD\in\grp {{SO}}(n)\:\text{ such that }\:\B{F}(\tI\otimes\tD)\K{\Phi}<0,
\end{equation}
and then $\forall n\neq 3$ Postulate FAITHE fails. For even $n\geq 2$ the situation is the same of
Proposition \ref{prop:O(n)} because the automorphism $\tD$ reversing the direction of every vector
is a rotation (around no axis\footnote{For example a $\pi$-rotation of a disk ($n=2$) is not around
  an axis of the disk.}). For odd $n$, in order to achieve a $\Psi$ such that $F(\Psi)<0$, it is
sufficient to take the automorphism $\tD$ corresponding to the rotation of the $n$-dimensional ball
around the $n$-th axis. The representative of $\tD$ is the diagonal matrix with $D_{ii}=-1$ for
$i=1,\ldots,n-1$ and $D_{nn}=D_{n+1n+1}=1$. Therefore the state
$\Psi=(\tI\otimes\tD)\Phi=-\sum_{i=1}^{n-1}(\lambda_i\otimes\lambda_i)
+\lambda_{n}\otimes\lambda_{n}+\lambda_{n+1}\otimes\lambda_{n+1}$ achieves $F(\Psi)<0$ for each odd
$n\geq 5$.\qed

\par The last two Propositions show that, among the probabilistic theories having as local system an
$(n)$spin-factor with $\grp{{SO}}(n)$ as group of physical automorphisms, it is possible to satisfy
Postulate FAITHE iff $n=3$. Therefore, according to Corollary \ref{cor:FAITH_tel_cor} and
Proposition \ref{prop:entsw} teleportation and uniqueness (modulo local automorphisms) of
purification at all levels can be satisfied. This is not surprising because the \emph{qubit} is
exactly the hidden quantum model (in the sense of Sec. \ref{s:hiddqm}) of the $(3)$spin-factor
probabilistic theory having $\grp{{SO}}(3)$ as physical automorphisms.

\section{Toy-theory 4: the classical world}
A probabilistic theory is said to be classical iff its local set of states $\Stset$ is a
\emph{simplex}. Including these theories in our probabilistic framework we can easily show how some
fundamental features of the classical theories arise from the simplex nature of $\Stset$.
Differently from the previous models the classical ones can be easily investigated on a generic
dimension.
\subsection{Probability simplex representation}
Consider a simplex set of states $\Stset$ with $\dim(\Stset)=n$ and denote as usual by $\mat
l=\{l_i\}$ and $\mat\lambda=\{\lambda_j\}$, with $i,j=1,\ldots,n+1$, the canonical basis for
$\Stset_\Reals$ and $\Cntset_\Reals$ as the same Euclidean space $\Reals^{n+1}$. The usual Bloch
representation---in which the deterministic effect corresponds to the vector
$\mat\lambda(e)=[0,\ldots,0,1]\in\Reals^{n+1}$---here becomes not convenient. A more convenient
representation of the simplex $\Stset$ is the so called \emph{probability simplex}, namely the
$n$-dimensional polyhedra whose $(n+1)$ vertices correspond to the canonical base vectors
$\{\lambda_i\}.$\footnote{Differently from the probabilistic models analysed until now, here the
  base vectors $\{\lambda_i\}\in\Stset_\Reals$ are true states of the classical theory.} Naturally
the cone of states $\Stset_+$ is the $\Reals^{n+1}$ positive orthant
\begin{equation}\label{e:CL_states}
\Stset_{+}=\Reals^{n+1}_{+}=\left\{\mat l(\omega)\in\Reals^{n+1}\;|\;\mat l(\omega)\succeq 0\right\},\quad
\Stset=\left\{\mat l(\omega)\in\Reals^{n+1}\;|\;\mat l(\omega)\cdot\mat 1=1\right\},
\end{equation}
where the symbol $\succeq$ denotes componentwise inequality\footnote{Componentwise or vector inequality in $\Reals^{n}$: $w\succeq v$ means $w_{i}\geq v_{i}$ for $i=1,\ldots,n$.} and $\mat 1$ denotes the vector $[1,\ldots,1]\in\Reals^{n+1}$. In this representation the system is pointedly self-dual and the cone and set of effects are respectively
\begin{equation}\label{e:CL_effects}
\Cntset_{+}=\Reals^{n+1}_{+}=\left\{\mat\lambda(a)\in\Reals^{n+1}\;|\;
\mat\lambda(a)\succeq 0\right\},\quad
\Cntset=\left\{\mat\lambda(a)\in\Reals^{n+1}\;|\;0\preceq\mat\lambda(a)\preceq 1\right\}.
\end{equation}  
The deterministic effect $e$, which must satisfy the condition $\omega(e)=1$ $\forall\omega\in\Stset$, and then $\lambda_i(e)=1$ $\forall i$, is now represented by the vector $\mat\lambda(e)=\mat 1\in\Reals^{n+1}$.
\par
\begin{figure}[ht]
 \begin{minipage}[c]{4.9cm}
  \centering
   \includegraphics[width=4cm]{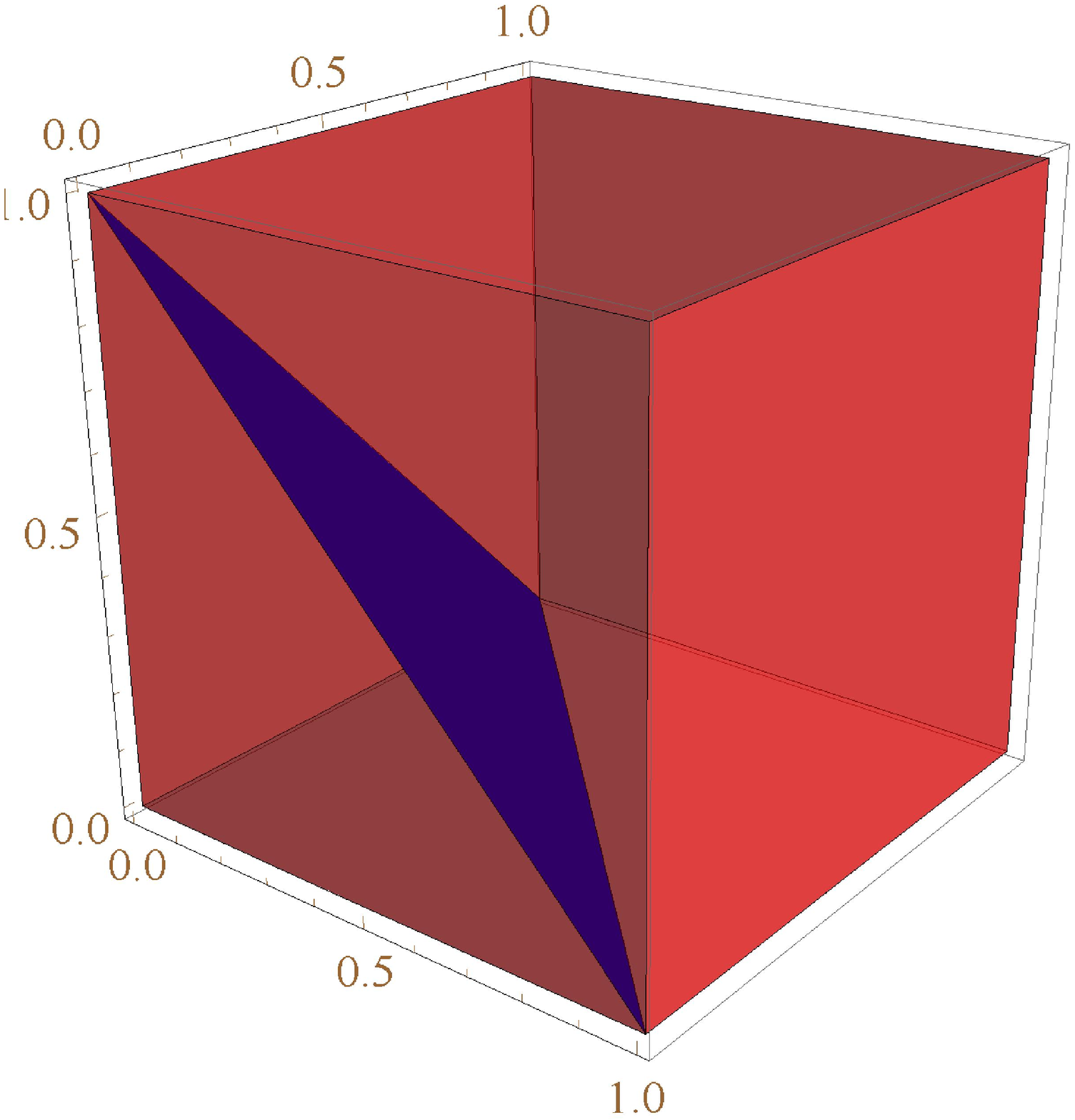}
 \end{minipage}
 \ \hspace{2mm} \hspace{2mm} \
 \begin{minipage}[c]{6.6cm}
   \centering
   \includegraphics[width=5cm]{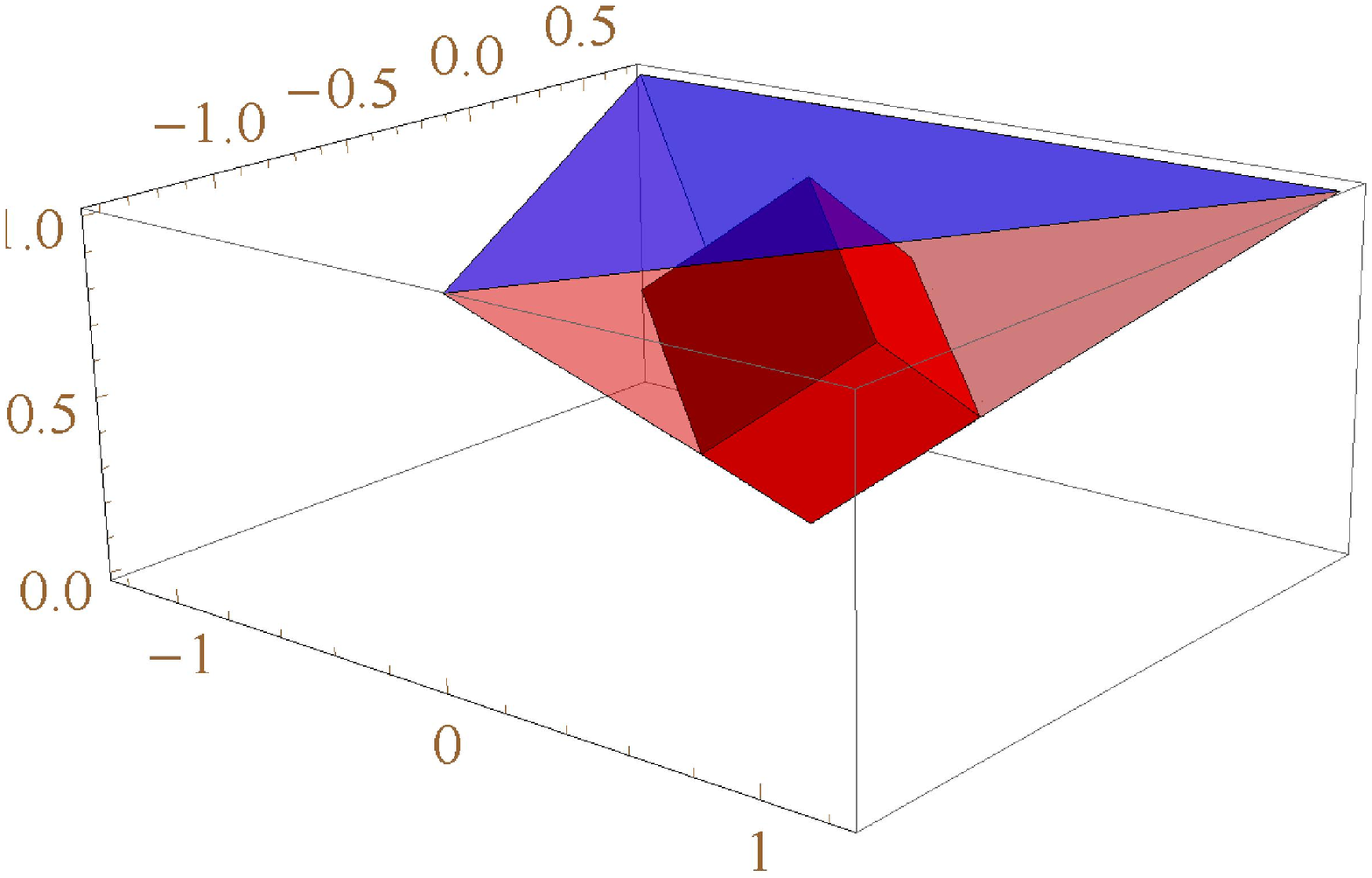}
 \end{minipage}
 \caption{{\bf Left figure:} Probabilistic simplex representation of the trit system. The triangle
   inside the cube represents the simplex of states $\Stset$ while the transparent cube is
   $\Cntset$. Both the cones $\Stset_+$ and $\Cntset_+$ coincide with $\Reals_+^{3}$.{\bf Right
     figure:} Bloch representation of the same trit system. The transparent cone represent both
   $\Stset_+$ and $\Cntset_+$. The triangle at the top is the simplex of states $\Stset$ while the
   convex of effects $\Cntset$ is the inside solid cube.}\label{fig:trit}
\end{figure}

 To clarify the situation we give a concrete representation of the classical theory with
$\dim(\Stset)=2$. The simplex in dimension 2 is a triangle and the corresponding system is called
{\bf trit}, a generalization of the {\bf bit} having a segment as simplex of states. In the left
Fig. \ref{fig:trit} we show the probabilistic simplex representation of the trit system according to
Eqs. (\ref{e:CL_states}) and (\ref{e:CL_effects}). For completeness in Fig. \ref{fig:trit}
the usual Bloch representation of the same system is also reported.

\subsection{Simplex structure consequences}The first consequence of the simplex nature of $\Stset$
is expressed in the following proposition.
\begin{proposition}\label{prop:simpl_struct}A probabilistic theory has a simplex as local convex set
  of states if and only if the bipartite set of states is a simplex too.
\end{proposition}
\Proof Let $\Stset$ be an $n$-dimensional simplex. We can denote by
$\omega_1,\omega_2,\ldots,\omega_{n+1}$ the vertices of $\Stset$. Then the set of functionals
$\{a_1,a_2,\ldots,a_n+1\}\in\Cntset_\Reals$ such that
\begin{equation}
a_i(\omega_j)=\delta_{ij}
\end{equation}
are vertices of $\Cntset$. Notice that in the probability simplex representation the vertices
$\Extr(\Stset)=\{\omega_1,\omega_2,\ldots,\omega_n+1\}$ coincide with the orthonormal basis
$\{\lambda_i\}$ for $\Stset_\Reals$. The cone of physical transformations $\Trnset_+$
($\dim(\Trnset_+)=(n+1)^2$) for a classical theory is the cone of positive maps, namely the maps
preserving the local cone of states $\Stset_+$. Then a map $\tA\in\Trnset_+$ if and only if
$\tA\omega\in\Stset_+$ $\forall\omega\in\Extr(\Stset)$, or, in the probabilistic simplex
representation, $\tA\lambda_i\in\Stset_+$ $\forall\lambda_i$. Being $\{\lambda_i\}$ the canonical
base, it follows that $\Trnset_+$ includes all the transformations represented by a $(n+1)\times(n+1)$
matrix with all non negative elements. Then in the probabilistic simplex representation the extremal
rays $\Erays(\Trnset_+)$ are generated by the $(n+1)^2$ matrices having an entry equal to one and
all the other entries equal to zero. In a generic representation these rays are the transformations
\begin{equation}\label{e:extr_tr_simp}
\gamma\omega_i\otimes a_j\qquad\forall i,j=1,\ldots,n+1,\;\forall\gamma\geq 0.
\end{equation}
where $\gamma$ is a multiplicative constant spanning the whole ray generated by the transformation
$\omega_i\otimes a_j$. These maps send the convex set $\Stset$ into an extremal ray of $\Stset_+$.
The preparationally faithful state of the theory $\Phi$ provides the isomorphisms
$\Cntset_+\simeq\Stset_+$ ($\Phi(\cdot,a_j)=\omega_j$) and $\Trnset_+\simeq\Stset_+(\rA\rA)$.
Remembering that a cone isomorphism preserve the cone structure, from the $(n+1)^2$ extremal rays of
$\Trnset_+$ in Eq. (\ref{e:extr_tr_simp}) we get the following $(n+1)^2$ extremal rays of
$\Stset_+(\rA\rA)$
\begin{equation}\label{e:extr_St_simp}
\gamma\omega_i\otimes\omega_j\qquad\forall i,j=1,\ldots,n+1,\gamma\geq 0.
\end{equation}
Then the only bipartite pure states of the theory are the $(n+1)^2$ factorized states
$\omega_i\otimes\omega_j$. In conclusion the bipartite set of states is a $((n+1)^2-1)$-dimensional
convex set having $(n+1)^2$ vertices, which is a simplex.

The opposite implication, if $\Stset(\rA\rA)$ is a simplex then $\Stset$ is a simplex too, is
trivial. Consider for example a $(n^2-1)$-dimensional bipartite simplex, then $\Stset(\rA\rA)$ has
only $n^2$ pure states. Naturally $\Stset$ cannot admit more than the $n$ vertices generating the
$n^2$ pure bipartite ones. Therefore $\Stset$ is a simplex.\qed This proposition has some
interesting corollaries which show the peculiarity of the classical theories with respect to the
other probabilistic theories. 
\begin{corollary}\label{cor:simplex1}The classical probabilistic theories are local.
\end{corollary}
\Proof A theory is said to be local if and only if it does not violate the CHSH inequality. The last
proposition shows that if the local set of states is a simplex then also the bipartite one is a
simplex and its vertices are factorized states. Then all the bipartite states are factorized
probability rules which do not allow violations of the CHSH inequality.\qed 

\bigskip In the following corollary we give a property of the set of local automorphisms for a
classical probabilistic theory.  The set of automorphism $\set{Aut}(\Stset)$ of an $n$-dimensional
simplex is the \emph{permutation group} $\grp S_{n+1}$, which contains the $(n+1)!$ different
permutations of the set $\Extr{\Stset}=\{\omega_1,\ldots,\omega_{n+1}\}$.
\begin{corollary}\label{cor:simplex2} The local automorphisms of a classical probabilistic theory
  cannot be extremal transformations.
\end{corollary}
\Proof A general element of $\set{Aut}(\Stset)=\grp S_{n+1}$ can be identified by a set of indexes
\begin{equation}
J=\{j_1,\ldots,j_{n+1}\},
\end{equation}
representing a permutation of the set $\{1,\ldots,n+1\}$. The automorphism associated to such
permutation is the map
\begin{equation}
\sum_{i=1,\ldots,n+1}\omega_{j_i}\otimes a_i,
\end{equation}
which is manifestly a convex combination of the extremal transformations $\omega_{j_i}\otimes a_i$
given in Eq. (\ref{e:extr_St_simp}) of Proposition \ref{prop:simpl_struct}.\footnote{In the case of
  the probability simplex representation the set of automorphisms are the $(n+1)\times(n+1)$
  permutation matrices which are manifestly combinations of the extremal ones, namely the
  $(n+1)\times(n+1)$ matrixes with an entries equal to one and the other entries equal to zero.}\qed

\begin{corollary}\label{cor:simplex3} The classical probabilistic theories do not satisfy Postulates PFAITH and PURIFY.
\end{corollary}
\Proof Notice that the identical transformation $\tI$ is a particular permutation $\tI\in\grp
S_{n+1}$ and then a local automorphism of the classical theory. According to Corollary
\ref{cor:simplex2} the identical transformation cannot be atomic. On the other hand we know from
Subsec. \ref{ss:PFAITH} that Postulate PFAITH implies the atomicity of $\tI$, whence it cannot be
satisfied. For the same reason also Postulate PURIFY does not hold. In fact, according to Lemma
\ref{lem:PURIFY_Id_Atomicity}, it implies atomicity of the identical transformation.\qed

It is not surprising that PFAITH fails. It assumes the existence of a pure preparationally faithful
state. On the other hand, as showed in Proposition \ref{prop:simpl_struct}, the only pure bipartite
states for a classical probabilistic theory are the factorized ones. These states obviously do not
achieve the isomorphism $\Stset_+\simeq\Cntset_+$ and then they are not preparationally faithful.
Therefore, a preparationally faithful state cannot be pure and PFAITH fails. Also the impossibility
of purifying a classical theory is almost obvious, since there are not enough bipartite pure states
to purify the continuous of internal points of the $n$-dimensional simplex $\Stset$. Precisely,
being the only bipartite pure states the $(n+1)^2$ factorization of the $(n+1)$ pure states of
$\Stset$, no mixed state admits purification. A similar problem is suffered by the extended
Popescu-Rohrlich model ( see Subsec. \ref{ss:PR_purification}) where no mixed state in $\Stset$,
apart from its center $\chi$, allows purification.
\par

\end{document}